\documentclass[pre,showpacs,notitlepage,twocolumn,superscriptaddress]{revtex4-1}

\pdfoutput=1
\usepackage{tabularx}
\usepackage{blindtext}
\usepackage{svg}
\usepackage{float}
\usepackage{mwe}
\usepackage{multirow}
\usepackage[colorlinks,linkcolor={blue},citecolor={blue},urlcolor={blue}]{hyperref}

\usepackage{graphicx}
\graphicspath{{Figures/}}
\usepackage{bm,amsmath}

\usepackage{latexsym}

\usepackage{verbatim}

\usepackage{xcolor}
\usepackage{subfigure}

\usepackage{gensymb}

\usepackage{soul}

\usepackage{tikz}

\renewcommand*\vec[1]{\mathbf{\bm{#1}}}
\begin{document}

\title[]{Active motion of tangentially-driven polymers in periodic array of obstacles}% 
% \title[]{Transport of active semiflexible polymers in ordered and disordered  media}% Force line breaks with \\
% Inertia-induced rigidity and enhanced dynamics in active polymers
%\thanks{Footnote to title of article.}
\author{Mohammad Fazelzadeh}
 %\altaffiliation[Also at ]{Physics Department, XYZ University.}%Lines break automatically or can be forced with \\
\affiliation{Institute of Physics,  University of Amsterdam, Amsterdam, The Netherlands
}%

\author{Ehsan Irani}
\affiliation{Institute for Theoretical Physics, Georg-August University of Göttingen, Friedrich-Hund Platz 1, 37077 Göttingen, Germany
}%

\author{Zahra Mokhtari}
\affiliation{Department of Mathematics and Computer Science, Freie Universität Berlin, Arnimallee 6, 14195 Berlin, Germany
}%

\author{Sara Jabbari-Farouji $^{*}$}
 %\altaffiliation[Also at ]{Physics Department, XYZ University.}%Lines break automatically or can be forced with \\
\affiliation{Institute of Physics,  University of Amsterdam, Amsterdam, The Netherlands
}%
 \email{Correspondence to: s.jabbarifarouji@uva.nl}
%\altaffiliation{}%Lines break automatically or can be forced with \
%\author{C. Author}
% \homepage{http://www.Second.institution.edu/\simCharlie.Author.}
%\affiliation{%Second institution and/or address%\\This line break forced% with \\}%

 \date{October 3, 2023}% It is always \today, today,
%This remarkable kinetics of MIPS stems from a competition between activity-induced accumulation of particles and inertia-induced suppression of clustering process.  In principle, our predictions can be tested by experiments with robotic worms. 
%\preprint{AIP/123-QED}

 %\date{\today}% It is always \today, today,
             %  but any date may be explicitly specified

%\pacs{ PACS}% PACS, the Physics and Astronomy
                             % Classification Scheme.
%http://hyperphysics.phy-astr.gsu.edu/hbase/oscda.html
  \begin{abstract}
  We computationally investigate the active transport of tangentially-driven polymers with varying degrees of flexibility and activity in two-dimensional square lattices of obstacles. Tight periodic confinement induces notable conformational changes and distinct modes of transport for flexible and stiff active filaments.  It leads to localization and caging  of flexible polymers inside the inter-obstacle pores, while promoting more elongated conformations and enhanced diffusion for stiff polymers at low to moderate activity levels. The migration of flexible active polymers occurs via hopping events, where they unfold to move from one cage to  another. In contrast,   
stiff chains travel mainly in straight paths within inter-obstacle channels, while occasionally changing their direction of motion. %and switching to another perpendicular channel.
Both the duration of caging and persistent directed migration within the channels decrease with increasing the activity level. 
As a consequence, at high active forces   polymers overcome confinement effects and transport within inter-obstacle pores as swiftly as those in free space. We explain the center of mass dynamics of semiflexible polymers in terms of active force and obstacle packing fraction by developing an  approximate analytical theory.
%Motion of flexible filaments consists of a sequence of  caging within pores and hopping events whereas motion of stiffer filaments primarily takes place via reptile-like migration within  the interobstacle channels and  occasional switching to a perpendicular channel. Under strong confinement and at a given activity, we observe a transition from localization to diffusion and channel switching upon increase of the chains stiffness.  

%In the flexible limit, the motion can be accounted for by the model of a continuous time random walk with a renewal process leading to unwinding of polymers and hopping to a new cage. The waiting time distribution is shown to develop heavy tails for decreasing stiffness, resulting in subdiffusive and ultimately caged behavior. We show that, depending on the degree of flexibility and strength of activity
  
  \end{abstract}

\maketitle
\section{Introduction}

%An important class of active  systems  consist of self-propelled units that use the energy from their surrounding environment to make directed motion~\cite{ActiveMatter}. Active particles can vary in several physical features, such as size, shape and their degree of flexibility. Examples range from  micron-sized entities such as motile cells~\cite{MotileCells}, self-propelled  biopolymers~\cite{biopolymers1,biopolymers2} and  active Janus colloids~\cite{ActiveJanus} to macroscopic worms~\cite{DEBLAIS,ANTOINEPhaseSeperation}, fish  and  birds~\cite{ActiveMatter}.

 Understanding the  dynamics of active particles within heterogeneous media,  subjected to intricate geometric constraints, has become an increasingly important research focus~\cite{Bechinger}.  This elevated interest arises due to the omnipresence of porous media in both natural settings, such as gels, tissues, and soils, and man-designed devices like array of micro-pillars in bio-technological applications. Gaining such insight is of relevance from both fundamental and applied perspectives.  
  From a fundamental viewpoint, we are interested in understanding  the impact of the heterogeneity of environments on the stochastic transport of active particles within complex media.
   From a practical standpoint, it helps us to unveil the movement and search strategies employed by living organisms in real-world environments. Furthermore, the knowledge acquired can be harnessed to pioneer innovative technological applications for controlling the motion of active agents within complex media.  For example,  smart self-propelled  carriers  can be used for cargo or drug delivery in heterogeneous media  or contamination removal in porous soil. 
%a wide range of biological, medical, and industrial situations.   

   In the last decade, both experimental and theoretical research efforts ~\cite{Bechinger,Chepizhko1,Chepizhko2,BacterialTrapping,Reichhardt_2014,Reichhardt_2018,Morin_2017,ElhamBacteria,Elham,PolydisperseABPO,JANUSPorous,Danne-M-van-Roon-Chiral,Roberto-Alonso-Matilla-OrderedMedia,mobilityvsmotility,ABP_periodic,Chamolly_2017,Chopra_22} have been directed at understanding the motion of active particles in heterogeneous porous media. The majority of studies so far have focused on the migration of active particles in disordered media~\cite{Bechinger,Chepizhko1,Chepizhko2,ElhamBacteria,Elham,BacterialTrapping,Reichhardt_2014,Reichhardt_2018,Morin_2017} as  most of movement environments for active particles in the nature are disordered. However,  patterned structures with periodic lattices are also found in the nature, for instance in  antibiofouling surfaces such as cicada wings~\cite{wing} and shark skins~\cite{shark_skin}. Moreover, a first approach to control the dynamics of active agents for technological applications relies on  designing  ordered heterogeneous media  by placing obstacles on  a substrate in periodic arrangements~\cite{AntoineSienceAdvances,Chopra_22}.

   Research in this direction includes studies of the motion of active particles and bacteria~\cite{ABP_periodic,mobilityvsmotility,Roberto-Alonso-Matilla-OrderedMedia,Danne-M-van-Roon-Chiral,Chamolly_2017,Chopra_22} in  periodic array of obstacles. These studies revealed that the periodic arrangement can enhance the persistent motion of  active particles at high enough activity levels~\cite{ABP_periodic} and more generally affect the transport efficiency of active particles, depending on the size of active particles and 
 the nature of their interactions with the structured medium.
      %However,  transport of active polymers in periodic array of obstacles has not been investigated before.
  %Although the effects of periodic array of obstacles on movement of  of active Brownian particles (ABP) and microswimmers~\cite{mobilityvsmotility,Roberto-Alonso-Matilla-OrderedMedia,Danne-M-van-Roon-Chiral} has been investigated
  %In the case of ABP, periodic arrangement induces directionality in the motion of ABP at late times and upon increase of the activity, persistent motion is enhanced and it increases with the size of the obstacles.
  While  the previous studies of active particles in periodic array of obstacles   focused on rigid or spherical swimmers, not much is known about the motion of flexible  elongated self-propelled particles,  such as active filaments in
  periodic environments. As a first step in this direction,  we investigate the motion of active polymers in a square lattice of obstacles. We note that although there have been several studies of active polymers in disordered porous media~\cite{Elham,PolydisperseABPO},  to our knowledge,  effects of a periodic arrangement of obstacles on conformation and dynamics of active filaments remain unexplored. We choose to focus on tangentially-driven active polymers in a 2D square lattice which are relevant for experiments of \emph{T.Tubifex} worms or E-coli bacteria  in ordered  two dimensional  array of pillars~\cite{AntoineSienceAdvances,Chopra_22}. 
  
  % Active semi-flexible Brownian polymers in polydisperse porous media are found to undergo shrinkage at low activity, while they swell at higher activities~\cite{PolydisperseABPO}. Simulation of active tangentially-driven polymers in two dimensional random porous media shows stiff polymers smoothly travel between obstacles while flexible ones get trapped between them~\cite{Elham}.

%\sara{please add the paper of Antoine in Science advances as well as the paper of frontiers on California worms.}
%in which there will be an energy cost to bend the back-bone and the orientation of active force on each monomer is parallel to the local tangent of the back-bone.
The key question that we address is under what conditions and how conformation and dynamics of active polymers are affected by the presence of ordered array of obstacles.  To this end, we study the conformational and dynamical features of polymers while varying their degree of flexibility, activity level and the porosity of the ordered medium via changing packing fraction of obstacles.
We find that the effect of periodic confinement becomes significant at high obstacle packing, where the pore size becomes smaller than the gyration radius of flexible polymers or persistence length of semiflexible polymers.  

Under tight confinement, flexible active polymers exhibit shrunken conformations which are predominantly caged in interobstacle pores  and their transport occurs via rare events where they unfold and hop to an adjacent cage.  %Increasing activity level enhances  the hopping probability  from one pore to another.
In contrast,  tight periodic confinement constrains   active filaments to travel in straight paths inside the periodic channels while occasionally bend and changing their directions. These two distinct modes of transport for flexible and stiff polymer  lead to different trends for the long-time diffusion of active polymers at low to moderate activity levels.
At low activities, tight periodic confinement decreases the long-time diffusion of flexible polymers, whereas for very stiff polymers it leads to enhancement of the long-time diffusion. Remarkably, for all the cases, at high activity levels, polymers exhibit a similar dynamics to those in free space, hence overcoming the effects imposed by tight geometrical constraints.

%In the former case, polymers get caged in interobstacle pores and their transport occurs via rare events where they unfold and hop to an adjacent cage.   In the latter case of stiffer polymers, tight periodic confinement constrains   active filaments to move in straight path within periodic channels accompanied by occasional U-turns where the filaments switch to another channel orthogonal to the initial one.

%Comparing the conformational and dynamical properties of different polymers, we find that the effects of the interactions with the obstacles become relevant at high degree of confinement and only for polymers with either very flexible or very stiff backbones. At high degree of confinement, flexible active polymers shrink in conformation and become caged between obstacles, while by increasing activity they can hop from one cage to the next one. In contrast, stiff chains in tight confinement, travel long distances ballistically while occasionally changing their direction. Our study shows how the combined effects of flexibility, activity and confinements enhance or hinder the transport of semi-flexible tangentially-driven active polymers in ordered media.

The remainder of this article is organized as follows. First, we introduce our simulation setup and the relevant set of dimensionless group of parameters. In section III, we investigate the effects of periodic arrays of obstacles on the conformation of active polymers under different situations.
%different conformational states of the polymers and examine the mean size and shape of them using the end-to-end distance and the bond correlation function. 
In section IV, we discuss distinct modes of transport for flexible and stiff polymers  under strong confinement.  We characterize the statistical features of caging events for flexible polymers and the channel switching events for stiff ones. In section V, we examine the effect of periodic confinement on dynamical properties of active polymers of varying degrees of activity and flexibility. Additionally, we  propose   analytical approach to rationalize  the dynamics of the center of the mass of  active chains under periodic confinement using suitable approximations.
%present our understanding on the different dynamical behaviour of the chains with low and high degree of flexibility. We study the orientational correlations of the chains as well as the mean squared displacements of the center of the mass of the polymers. We also propose an analytical solution for the motion of the center of the mass of chains using justified approximations. 
Finally, we summarize our most important findings and our concluding remarks in section VI.
 
 %\sara{Your outline is too detailed and technical. Keep this in mind for writing of your next article.}

%https://iopscience.iop.org/article/10.1088/1367-2630/aa8d5e/pdf
%https://iopscience.iop.org/article/10.1088/1367-2630/ac7d00/pdf
%https://link.springer.com/article/10.1140/epje/i2019-11826-7
\section{Simulations Details}
\subsection{Simulation model for active polymers  and their interactions with obstacles} 
\begin{figure}[t]
    \centering
    \begin{tikzpicture}
            \draw (0,0) node[inner sep=0]{\includegraphics[width=1\linewidth ]{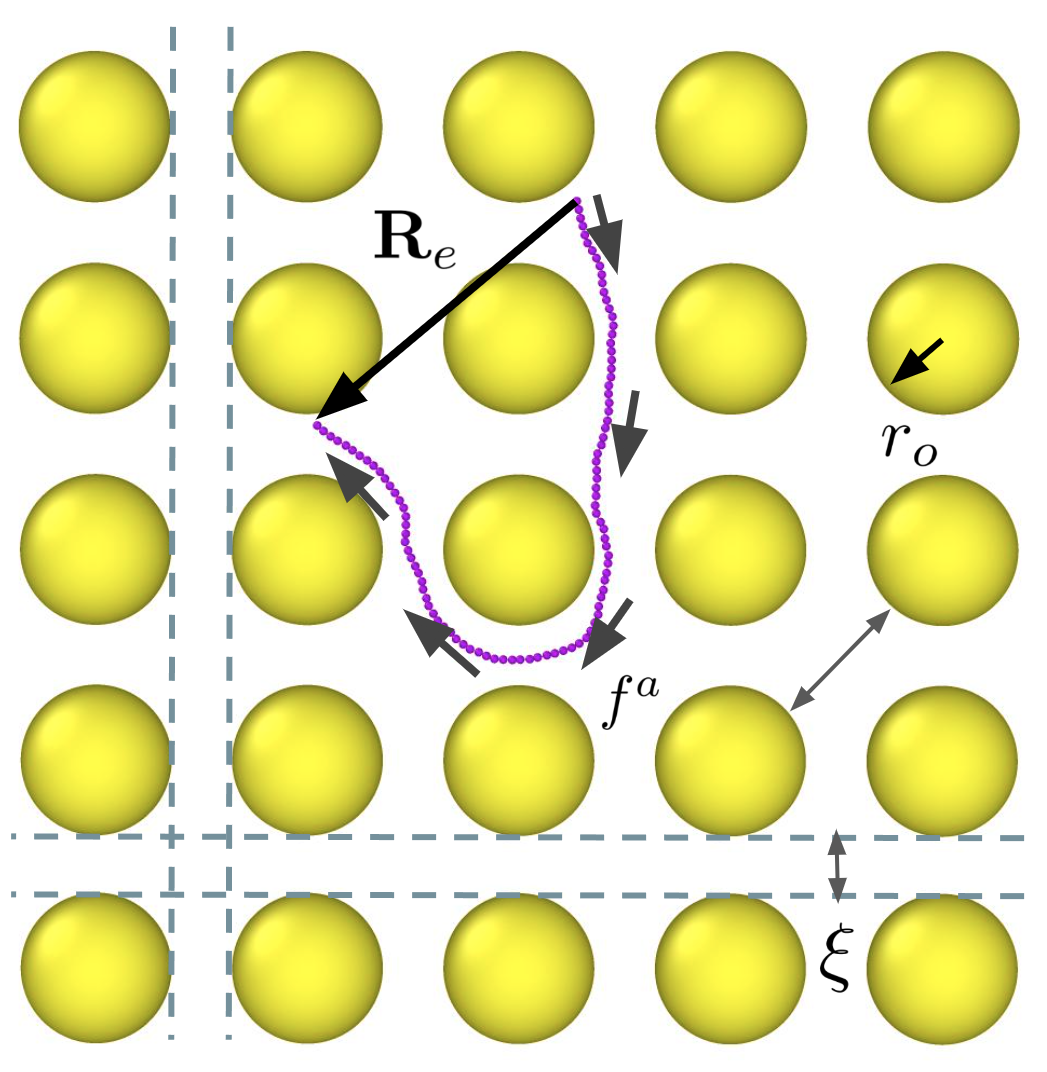}};
            \draw {(1.9,-1.03) node {\Large{$d_{cage}$}}};
            
    \end{tikzpicture}
    \caption{Schematic of the active tangentially-driven polymer in the ordered array of obstacles, showing the end-to-end vector $\vec{R}_e=\vec{r}_{N} -\vec{r}_{1}$, the radius of obstacles $r_o$, the pore size $d_{cage}$ and the horizontal/vertical channels of width $\xi$. }
    \label{fig:1}
\end{figure}

In order to  study the motion of semiflexible active polymers in ordered heterogeneous media,  % Better discuss this in introduction that why polymers are phantom, what is the experimental setup? how the worms have 3d freedom in 2d surface. 
we implement the tangentially-driven polymer model~\cite{IseleTangential} in a 2D square lattice of circular obstacles, see Fig.~\ref{fig:1}. 
In experiments~\cite{AntoineSienceAdvances,Chopra_22}, 2D projection of 3D active filaments around cylindrical pillars  are observed, which implies that polymer can cross itself. To mimic this situation, we discard excluded volume interactions between monomers and we consider
 a phantom active polymer model of $N$ monomers. The motion of each monomer 
  is governed by the overdamped Langevin dynamics and is given by
  \begin{equation}
    \gamma \dot{\vec{r}}_i = - \sum_j \nabla_{\vec{r}_i} U+\vec{f}^a_{i}+\vec{f}^{r}_{i},
    \label{eq:brownian}
\end{equation}
where $\vec{r}_{i}$ is the position of the $i$th monomer, the dot denotes the derivative with respect to time and $\gamma$ is the friction coefficient between the bead and its surrounding medium. 

The potential energy $U$ of each monomer includes three different contributions. The 
first one is the harmonic spring potential $U_{\text{harmonic}}(r)=(k_s/2)(r-\ell)^2$, with equilibrium length $\ell$ and spring stiffness $k_s$ between adjacent monomers. The Second part is the bending potential between each two neighbouring bonds $U_{\text{bend}}(\theta_i)=\kappa(1-cos\theta_i)$, where $\theta_i$ denotes the angle between two consequent  bonds intersecting at bead $i$ defined as $\theta_i= \cos^{-1} (\widehat{\vec{t}}_{i,i+1} \cdot \widehat{\vec{t}}_{i-1,i})$ with  $\widehat{\vec{t}}_{i,i+1}= \vec{r}_{i,i+1}/|\vec{r}_{i,i+1}|$ and $\vec{r}_{i,i+1}=\vec{r}_{i+1}-\vec{r}_{i}$.  Here, $\kappa$ is the bending stiffness and determines the intrinsic degree of flexibility of a polymer. Finally, the third contribution accounts for the excluded volume interactions between each bead and its surrounding obstacles. They are modelled by the short-ranged Weeks-Chandler-Andersen (WCA) potential~\cite{WCA}
$ U_{\text{excl}}(r)=4  \epsilon \left[  (\frac{\sigma/2+r_o}{r})^{12} -(\frac{\sigma/2+r_o}{r})^6+\frac{1}{4}\right]$
for $ r< r_c=2^{1/6} (\sigma/2+r_{o})$, where $\epsilon$ is the strength of the potential and has unit of energy, $\sigma$ is the diameter of the beads and $r_{o}$ is the radius of obstacles. The WCA potential is zero for interaction distances larger than the cutoff length $r_c$.\\

 The  active force on each bead, except for the end monomers, is given by:
 $\vec{f}^a_i=\frac{f^a}{2 \ell } (\vec{r}_{i-1,i}+\vec{r}_{i,i+1})$. The active force on the tail monomer is given by $\vec{f}^a_1=\frac{f^a}{2\ell } \vec{r}_{1,2}$  and for the head monomer is $\vec{f}^a_N=\frac{f^a}{2\ell } \vec{r}_{N-1,N}$.   For this model, the total active force  on each polymer is proportional to its end-to-end vector $\vec{F}^a(t) = f^a \vec{R}_{e}(t)/\ell$.  The random force is chosen as a white noise of zero mean and has the correlation $\langle \vec{f}^r_i(t) \cdot \vec{f}^r_j(t') \rangle=4 D_0 \gamma^2 \delta_{ij} \delta(t-t')$. To keep our formulation general, we do not associate random force  necessarily with thermal fluctuations, but it can also be of biological origin. The persistence length of a  2D passive ideal polymer in free space can be determined in terms of its bending stiffness and the strength of random force correlation as $\ell_p^0=2 \kappa \sigma/D_0 \gamma$~\cite{Binder_2020}.

  We use a fixed number of $N_o$ obstacles arranged in a square lattice in an $L\times L$ simulation box with periodic boundary conditions. To change the degree of confinement, we vary the packing fraction of the medium defined as the fraction of the surface occupied by the obstacles to the total area of the box $\phi=N_o \pi r_o^2/L^2$. The width of horizontal/vertical channels    $\xi$ is given by the free space between two neighbouring obstacles as shown in Fig.~\ref{fig:1} and it is obtained as $\xi=r_o(\sqrt{\pi/\phi}-2)$.  We also define the approximate cage diameter (pore size) as $d_{cage} = r_{o}(\sqrt{2\pi/\phi}-2)$, see Fig.~\ref{fig:1}. For flexible polymers, we can use the ratio of mean polymer size to the cage diameter as a measure of confinement strength, whereas for semiflexible polymers with $\kappa \gg 1$, the  ratio of the persistence length to the channel width  $\ell_p^0/\xi$ provides a good measure of confinement degree.

\subsection{Simulation parameters}
We choose $l_u=\sigma$, $E_u=\epsilon$ and $\tau_u=\gamma \sigma^2/\epsilon$ with $\gamma=1$ as the units of length, energy and time. From here on, we express quantities in dimensionless units. We subsequently fix $\ell=1$, $r_o=8.7$ and the diffusion coefficient $D_0=1$. The spring constants are chosen very stiff   $k_s  \gg f^a/\ell$,  to ensure that the mean bond-length and polymer contour length remain almost constant during simulations. We focus on the chain length $N=100$ and investigate the effects of periodic confinement on the conformational and the dynamical properties of active polymers of varying activity strengths  $ 0.001\le f^{a} \le 10$ and bending rigidity values $\kappa \in \{1,10,100\} \;\epsilon$. 
 We examine the conformation and dynamics of active polymers  under moderate   ($\phi=0.2$) and strong  ($\phi=0.6$) confinement leading to channel widths  $\xi \in \{17,2.5\}\; \sigma$, respectively and compare them with those of active polymers in free space ($\phi=0.0$).
%We also specifically include simulation of polymers in packing fraction of  $\phi=0.6$ with $\kappa = 1$ and $f^{a}\in \{0.3,0.4,2.0,5.0\}\; \epsilon/\sigma $ and polymers with  $\kappa = 100$ and $f^{a}\in \{0.005,0.01\}\; \epsilon/\sigma $   for a better understanding of the behaviour of the polymers. 
To analyze the effects of periodic arrays of obstacles on conformation and dynamics of the polymers, we studied ensemble-averaged conformational and dynamical properties of active filaments, where $\langle...\rangle$ indicates ensemble average over 120 independent simulation runs carried over a time span at least 5 times larger than the relaxation time of the end-to-end vector, for details please see supplemental material (SM~\cite{SI}, section I).

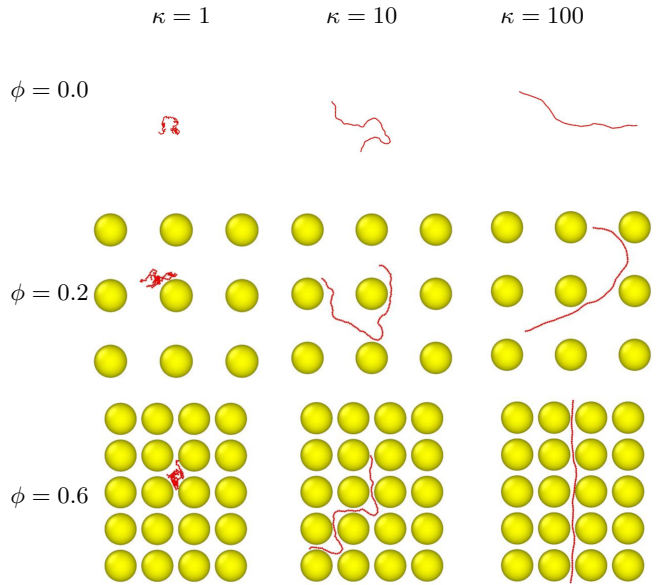
\begin{figure}[t]
    \centering
    \begin{tikzpicture}
            \draw (0,0) node[inner sep=0]{\includegraphics[width=0.9\linewidth ]{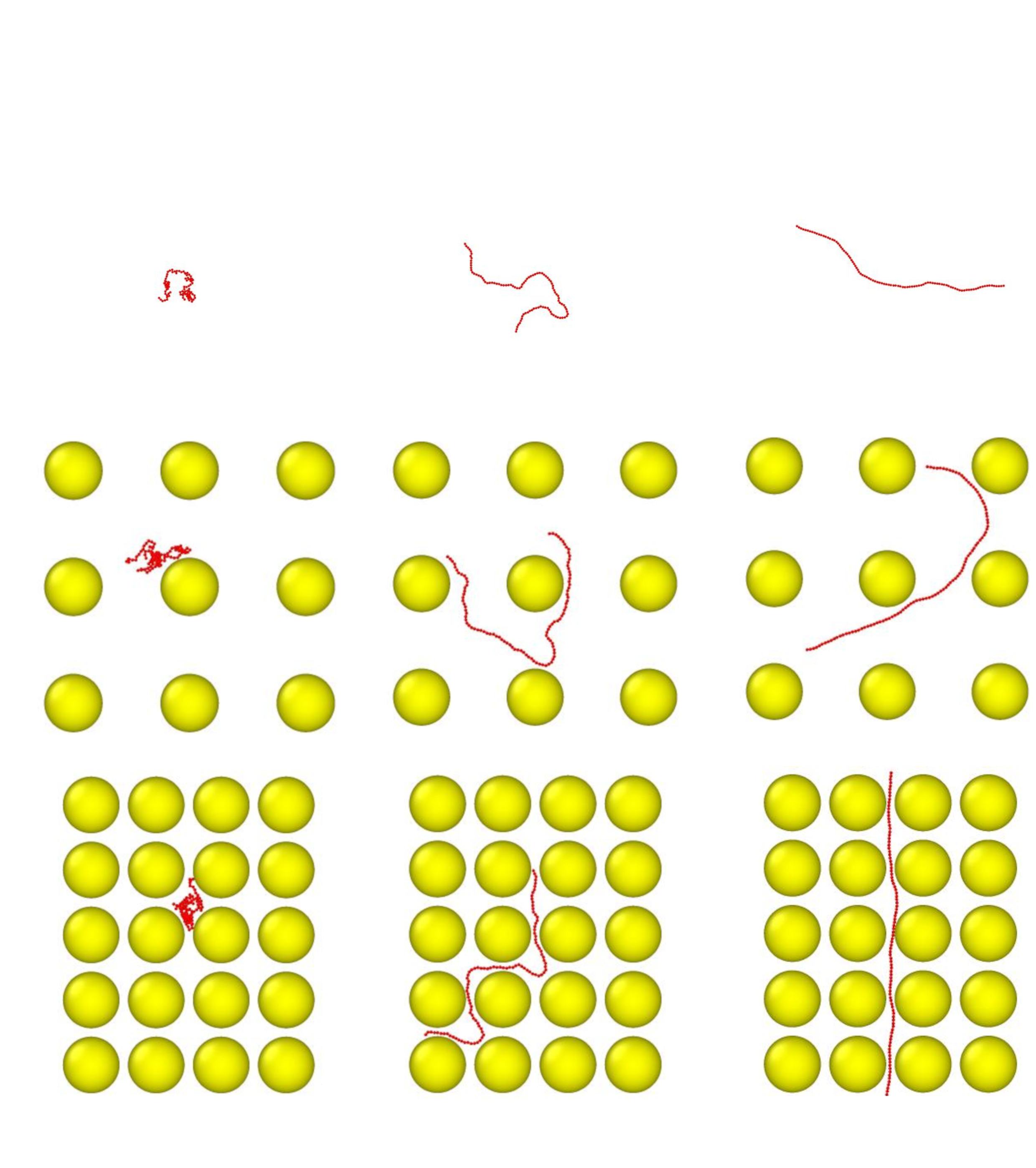}};
            
            %\draw {(-6,0.2) node {\includegraphics[width=0.40\linewidth {Figures/Fig0.1.png}}};
            \draw {(-4.15,2.7) node {$\phi=0.0$}};
            \draw {(-4.15,0.) node {$\phi=0.2$}};
            \draw {(-4.15,-2.7) node {$\phi=0.6$}};
            \draw {(-2.4,3.7) node {$\kappa=1$}};
            \draw {(0.,3.7) node {$\kappa=10$}};
            \draw {(2.4,3.7) node {$\kappa=100$}};
            %\draw {(-2.7,0) node {\rule[1pt]{1pt}{200pt}}};
            %\draw {(-7.8,3.7) node {$\kappa=1$}};
            %\draw {(-6.1,3.7) node {$\kappa=10$}};
            %\draw {(-4.,3.7) node {$\kappa=100$}};
            %\draw {(0.8,4.5) node {{\normalsize$f^a=1$}}};
            %\draw {(-6.1,4.5) node {{\normalsize$f^a=0.1$}}};
    \end{tikzpicture}
    \caption{Snapshots of an active chain with $f^a=1$ in various combination of $\phi$ and $\kappa$ . See videos S1-S9 at SM for temporal  evolution of an active polymer conformation moving through the obstacles. }
    
    \label{fig:SnapsORD}
\end{figure}

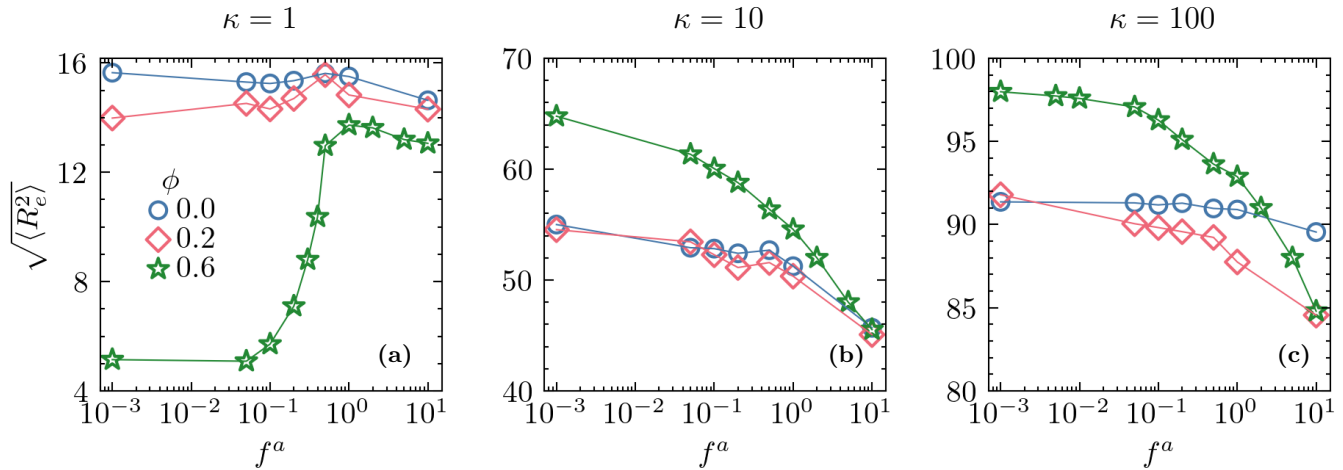
\begin{figure*}[t]
\vspace{1cm}
\centering
    \begin{tikzpicture}
            \draw (0,0) node[inner sep=0]{\includegraphics[width=\linewidth ]{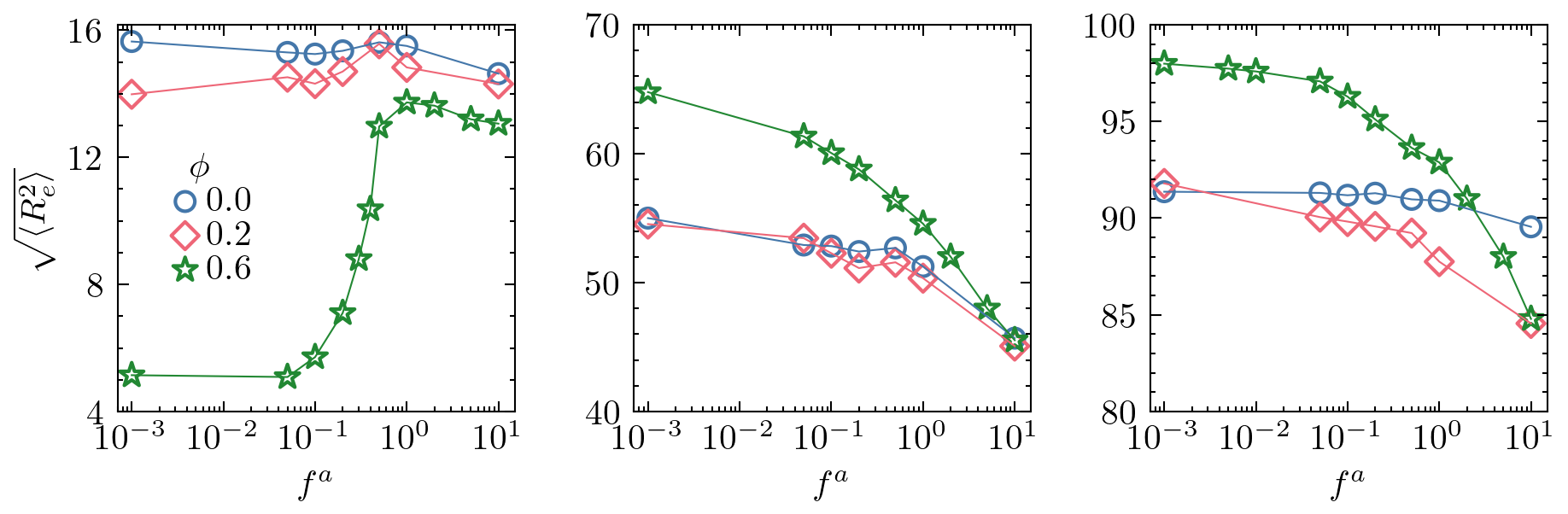}};
            \draw (-3.7,-1.3) node {\textbf{(a)}};
            \draw (2.3,-1.3) node {\textbf{(b)}};    
            \draw (8.25,-1.3) node {\textbf{(c)}};
            \draw (-5.5,3.2) node {{\large$\kappa=1$}};
            \draw (0.6,3.2) node {{\large$\kappa=10$}};
            \draw (6.5,3.2) node {{\large$\kappa=100$}};
    \end{tikzpicture}
    \caption{The root of mean-squared end-to-end distance $\langle R_e \rangle$ versus active force $f^a$ at  packing fractions $\phi=0,0.2$ and 0.6 plotted for bending rigidity values (a) $\kappa=1$, (b) $\kappa=10$  and (c) $\kappa=100$.} % \sara{Please add minor ticks for x-axis.Update panels (b) and (c) once you have more dara for larger $f^a$.}}
    \label{fig:ReORD}
\end{figure*}

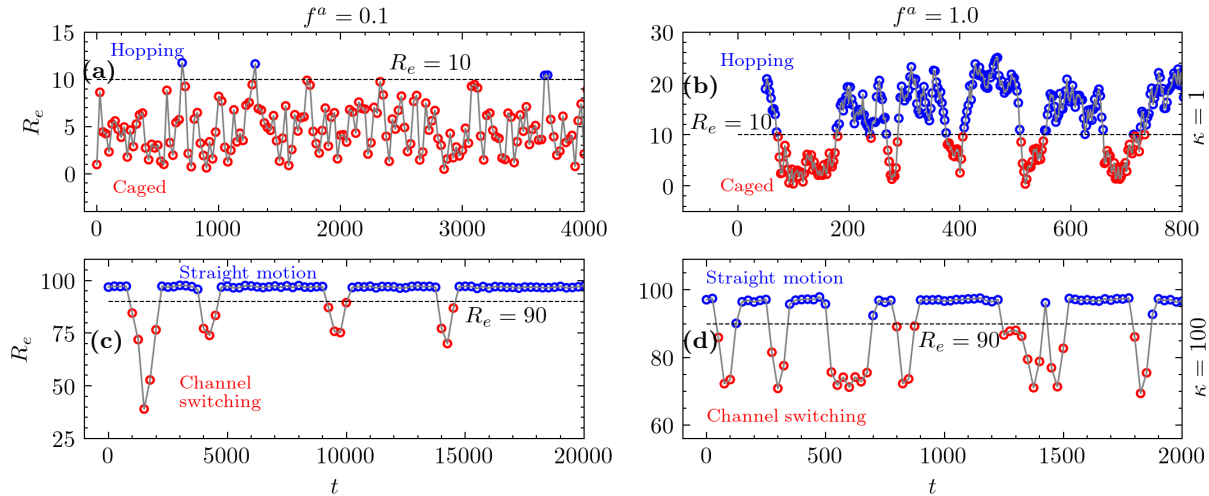
\begin{figure*}[t]
    \centering
    \begin{tikzpicture}
            \draw (0,0) node[inner sep=0]{\includegraphics[width=0.9\linewidth ]{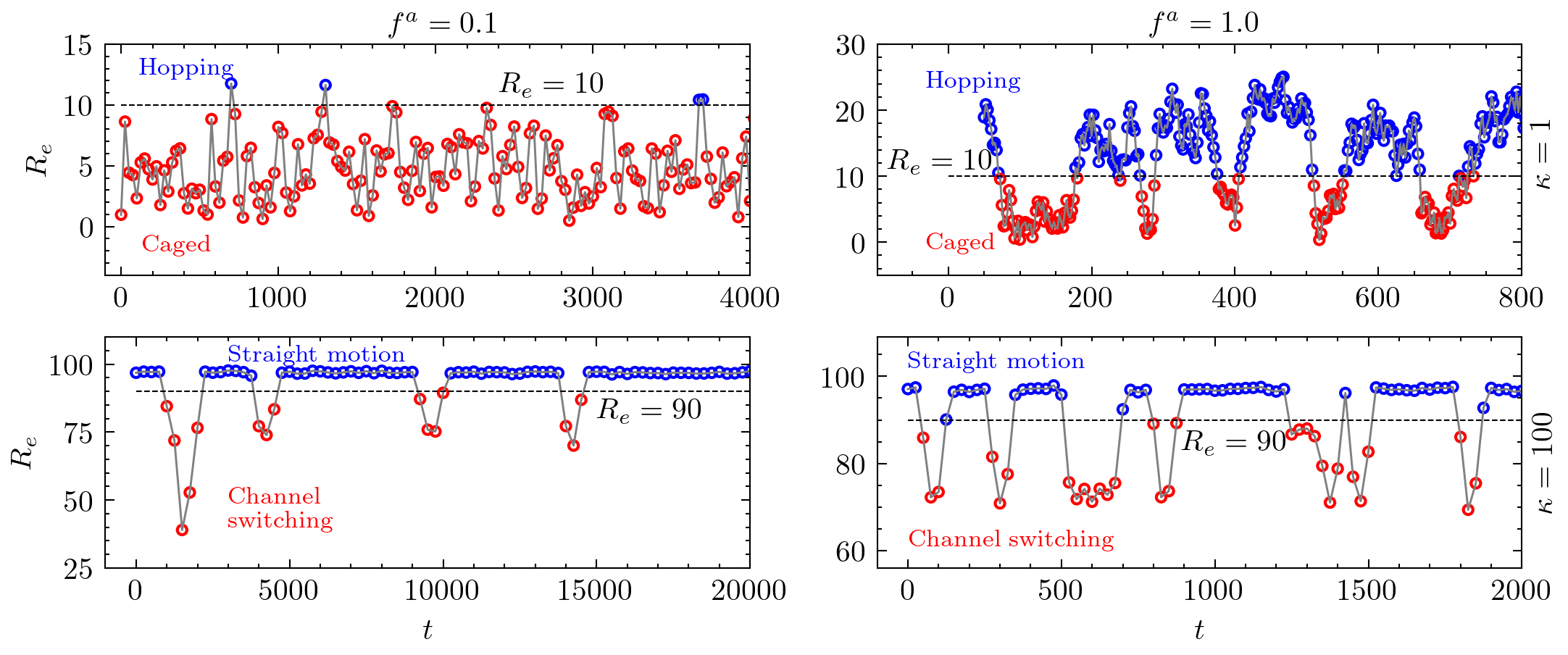}};
            \draw (-6.8,2.4) node {\textbf{(a)}};
            \draw (-6.7,-1.2) node {\textbf{(c)}};
            \draw (+1.2,2.2) node {\textbf{(b)}};
            \draw (1.2,-1.2) node {\textbf{(d)}};
            
    \end{tikzpicture}
    \caption{End-to-end distance of a single polymer  under strong confinement   $\phi=0.6$, with a) $\kappa=1$ and $f^a=0.1$, b) $\kappa=1$ and $f^a=1.0$, c) $\kappa=100$ and $f^a=0.1$ and d) $\kappa=100$ and $f^a=1.0$. The dashed lines show the threshold on $R_e^*$, above which chains are hopping or having straight motion depending on their stiffness. (a) and (b) show the characterization of hopping and caging events for flexible polymers. (c) and (d) show the characterization of straight motion through inter-obstacle channels and channel switching events for stiffer chains. }
    \label{fig:Re_trapp}
\end{figure*}

\begin{figure}[t]
    \centering
    \begin{tikzpicture}
            \draw (0,0) node[inner sep=0]{\includegraphics[width=\linewidth ]{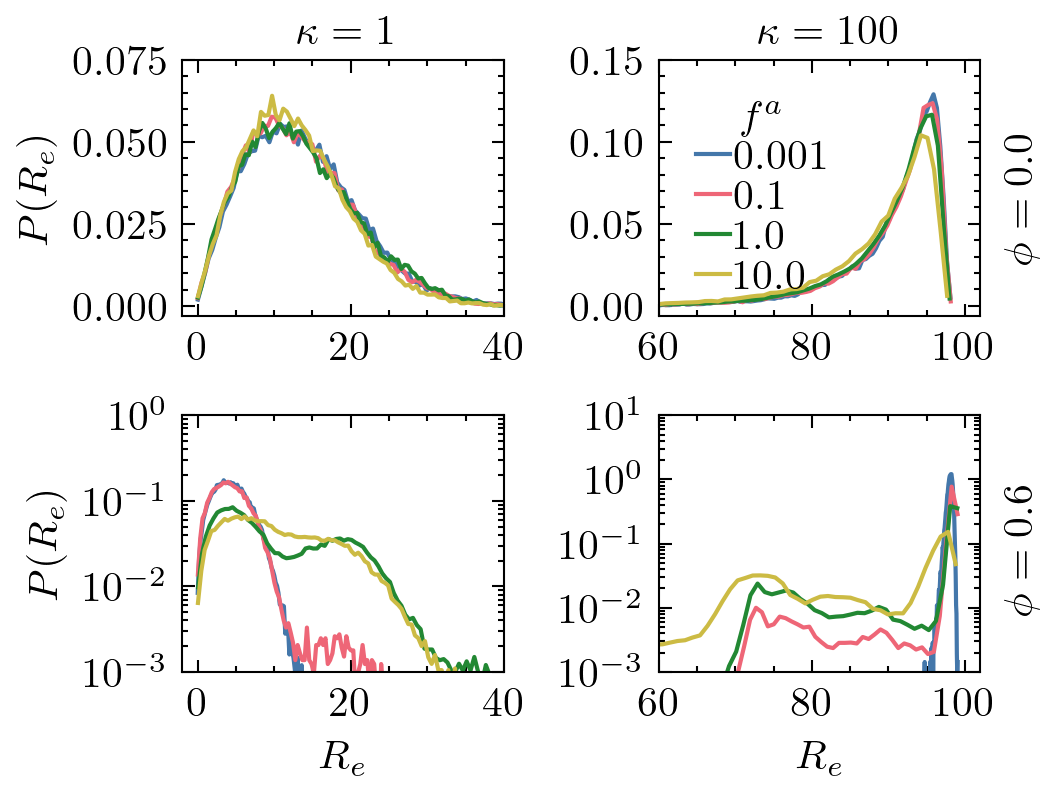}};
            \draw (-2.55,2.4) node {\textbf{(a)}};
            \draw (1.4,2.4) node {\textbf{(b)}};
            \draw (-2.55,-0.5) node {\textbf{(c)}};
            \draw (1.4,-0.5) node {\textbf{(d)}};
            
    \end{tikzpicture}
    \caption{The probability distribution function of end-to-end distance at different activity levels for polymers moving in free space {\it i.e.}, $\phi=0$ a) $\kappa=1$ and b) $\kappa=100$ and $\phi=0.0$ and for polymers under strong confinement  with $\phi=0.6$ c) $\kappa=1$ and d) $\kappa=100$. }
    \label{fig:PDFRe}
\end{figure}

%\section{Active polymers in ordered media}
 
\section{Conformational  properties }
 We begin by visually inspecting the effect of confinement degree  on chain conformation. In Fig.~\ref{fig:SnapsORD}, we present snapshots of active polymers with $f^a=1.0$ for various combinations of bending stiffness and packing fraction values. The overall trend that we observe for flexible chains ($\kappa=1$) is that by increasing the degree of confinement, at the highest studied packing fraction  
 ($\phi=0.6$) the polymers become localized in the cages of their adjacent obstacles for the majority of time. As a result, their  mean size shrinks. In contrast, for stiffer chains with $\kappa \geq10$, under strong confinement at $\phi=0.6$, the chains conformation becomes more extended and anisotropic as the persistence length becomes larger than the channel width.\\
 
%\subsection{Global chain conformation}
 
 \subsection{End-to-end distance}
  To quantify the effect of confinement degree on mean  conformation of active polymers, we examine the root  of mean-squared end-to-end distance $\sqrt{\langle R_e^2\rangle}=\sqrt{\langle |\vec{r}_N-\vec{r}_1|^2\rangle}$. Fig.~\ref{fig:ReORD} shows $\sqrt{\langle R_e^2\rangle}$ against activity for different bending stiffness values and packing fractions. We first note that in free space,  $\sqrt{\langle R_e^2\rangle}$ has a weak dependence on active force for all degrees of flexibility. It decreases very little upon increase of $f^a$. 
  Upon introducing periodic obstacles, for low  density of obstacles at $\phi=0.2$, $\sqrt{\langle R_e^2\rangle}$ decreases only slightly for all $\kappa$ values, the exact amount of which depends on the degree of activity and flexibility.  However,  for tight confinement at $\phi=0.6$, we observe a remarkable change of $\sqrt{\langle R_e^2\rangle}$ for all degrees of flexibility, albeit in contrasting trends for flexible and stiff polymers.

   For Flexible chains with $\kappa=1$, 
   %there is no considerable difference in $R_e$ of free chains and those with $\phi=0.2$ among different activities. 
  at $\phi=0.6$, chains of low activity $f^a \le 0.1$ have singnificantly smaller mean end-to-end distance than the free polymers with identical active force, whereas for $f^a> 0.1$ we observe a rather sharp increase of $\sqrt{\langle R_e^2\rangle}$   reaching a maximum at $f^a=1$   beyond  which  its values becomes almost constant being only  a little smaller than the $\sqrt{\langle R_e^2\rangle}$ of free polymers. Visual inspection of movement of flexible polymers with low levels of activity through the obstacles , see  Fig.~\ref{fig:SnapsORD} and video S10 in SM~\cite{SI}, reveals that in tight confinement, they remain localized within interobstacle pores,  referred to as cages, most of the time and only occasionally  unwind and hop to
  one of their adjacent cages. Figs. \ref{fig:Re_trapp}(a) and (b) depict  temporal evolution of end-to-end distance of an individual active polymer with $f^a=0.1$ and 1. When a chain is caged its end-to-end distances is small, but while hopping it has a more extended conformation and its end-to-end distance is larger.  
  We note that with increasing the activity level, the chains become enabled to escape their cages and travel a short distance to one of their four neighboring cages, increasing frequency of the hopping events.   %the time spent within a cage decreases and  frequency of hopping events increases. 
  For instance, at $f^a=1$ the chains hop more frequently from one cage to another, compare Figs. \ref{fig:Re_trapp}(a) and (b) and see video S11 in SM~\cite{SI}.  As a result  of their  increased hopping frequency, the  average end-to-end distance of highly active chains increases. \\

 As mentioned earlier, we observe a remarkably different trend for $\sqrt{\langle R_e^2\rangle}$ of stiffer chains.  As presented  in Figs.~\ref{fig:ReORD} (b) and (c), for $\kappa=10$ and $100$, tight confinement results in a larger end-to-end distance at all levels of activity. However, degree of extension of polymers decreases with increase of activity level. 
 At packing fraction of $\phi=0.6$, semiflexible active chains with $\ell_p^0=20$ and 200 ($\kappa=10$ and 100)  tend to travel through inter-obstacle horizontal/vertical channels as   their persistence lengths are much larger than the channel width $\xi=2.5$. Nonetheless, active polymers occasionally bend and turn  into an adjacent perpendicular channel, see Fig.~\ref{fig:SnapsORD} and video S9 in SM~\cite{SI}. We refer to this behavior as channel switching. During a channel switching event,  a  polymer bends and its end-to-end distance decreases,  see Fig.~\ref{fig:Re_trapp}(c) and (d) showing end-to-end distance of a stiff active polymer with $\kappa=100$ for $f^a=0.1$ and 1 versus time. 
 The folding of backbone of a polymer during a channel switching event results in higher bending energy compared to straight conformation, which is more costly for stiffer chains. However,  sufficiently large active forces can overcome the bending energy barriers for a channel switching event. As a result, upon increase of $f^a$, stiff chains  switch their channels more frequently,  see Fig.~\ref{fig:Re_trapp}(c) and (d)  explaining the decline of $\sqrt{\langle R_e^2\rangle}$ of confined polymers with active force. 
 Increase of the end-to-end distance in tight confinement is more pronounced for stiffer chains, for which channel switching events occur less frequently due to the higher costs of bending.  
 %We also note that  for a given level of activity the stiffer the chains, the higher the costs of bending, hence channel switching events occur less frequently for very stiff chains.  
  %make less number of turns and keep their traveling direction mostly in direction of channel, whereas for less stiff chains channel switching events occur more frequently. 

 To quantify our observation of different modes of chain conformation resulting from interaction with obstacles, we extract the probability distribution function (PDF) of the end-to-end distance $P(R_e)$ presented in Fig.~\ref{fig:PDFRe}. For active chains in free space, see Fig.~\ref{fig:PDFRe} (a) and (b), $P(R_e)$ displays only a single peak, the value of which is primarily determined by the stiffness of the polymers and weakly depends on the active force.
  %regardless of activity the values of $R_e$ are distributed around one peak value depending on the stiffness of the polymers.
 In contrast, under strong confinement at $\phi=0.6$, the PDFs of end-to-end distance becomes broader. At sufficiently high activity levels, we 
  distinguish two distinct peaks in the PDFs for both flexible and stiff polymers, verifying two  different conformational modes of active chains confined within periodic channels. For active flexible chains with $f^a \geq 0.1$, the two peaks correspond to conformations in caged  and hopping modes, respectively. For stiff chains with $f^a \geq 0.1$ the sharp peaks at  large $R_e$ correspond to elongated polymer conformations   traveling straight within the inter-obstacle channels whereas the broader peaks at smaller  $R_e$ represent  the chain  conformations in channel switching mode.

 %On panel (c) for flexible chains in tight confinement, when the activity is high ($f^a \geq 1.0$), $P(R_e)$ has a peak corresponding to a smaller conformation with $R_e\approx 5$ and another one corresponding a larger conformation with $R_e\approx 20$. The peak at shorter $R_e$ represents the flexible chains that are caged, whereas the peak at larger $R_e$ shows the chains that are hopping from one cage to a neighbouring one. Similarly, on panel (d) for stiff chains in tight confinement, the sharp peak at $R_e\approx 97$ shows the chains that are traveling in straight inter-obstacle channels. The distribution in the range $60 \lesssim R_e \lesssim 90$ represents the chains that are in the middle of channel switching.

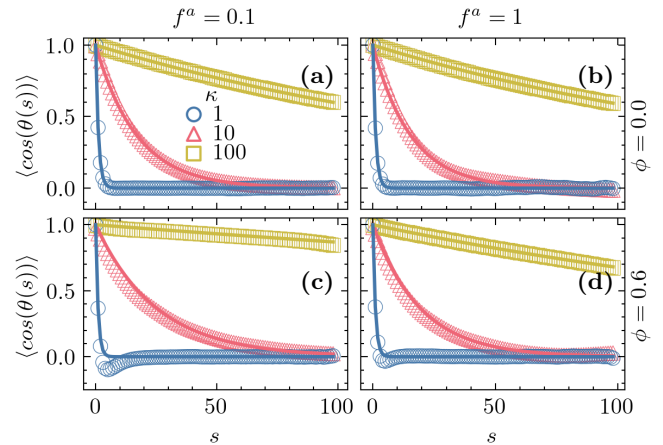
\begin{figure}[t]
    \centering
    \begin{tikzpicture}
            \draw (0,0) node[inner sep=0]{\includegraphics[width=\linewidth ]{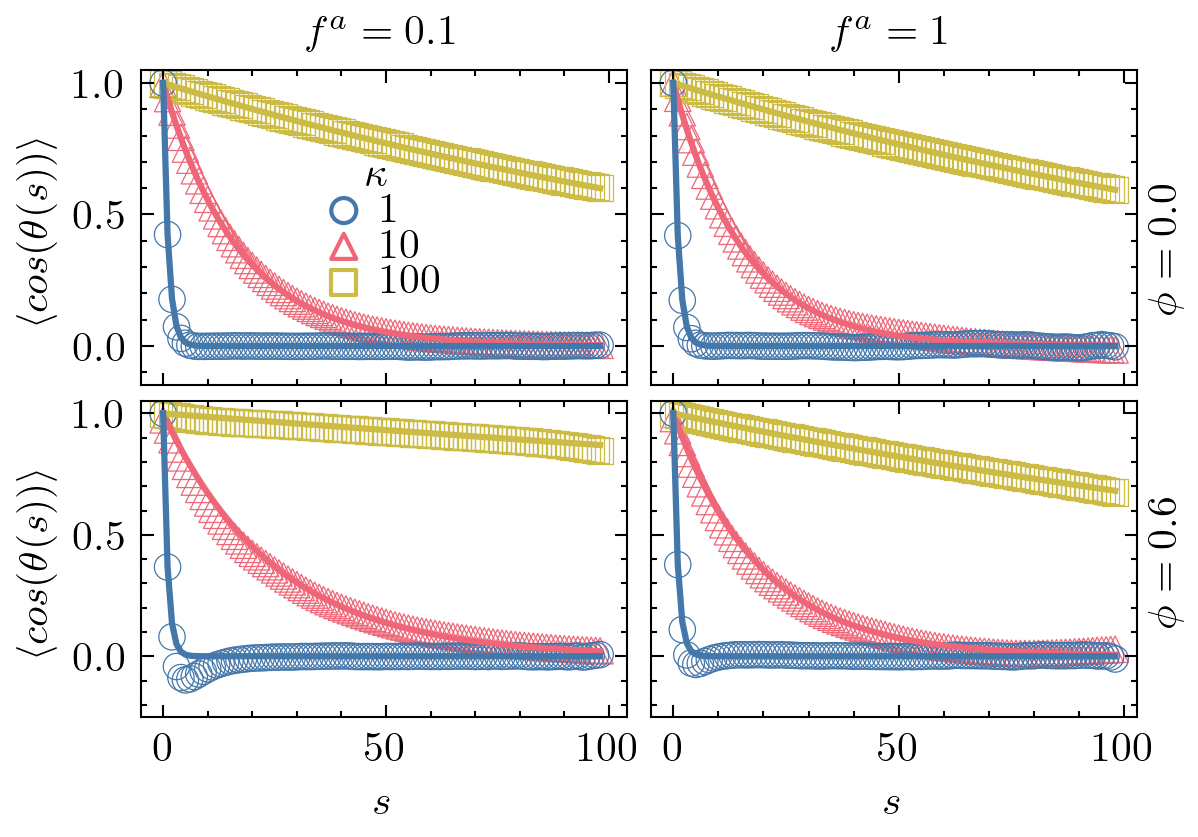}};
            \draw (-0.2,2.) node {\textbf{(a)}};
            \draw (3.5,2.) node {\textbf{(b)}};
            \draw (-0.2,-0.75) node {\textbf{(c)}};
            \draw (3.5,-0.75) node {\textbf{(d)}};
            
    \end{tikzpicture}
    \caption{The bond-bond correlation functions $\cos \theta(s)$ for various bending rigidity $\kappa$ values  for free chains with active force a) $f^a=0.1$ and b) $f^a=1$ and at packing fraction of $\phi=0.6$ for active force c) $f^a=0.1$ and d) $f^a=1$. The solid lines show the exponential fits to the data in the range $1\leq \langle cos(\theta(s)) \rangle \leq e^{-1}$.}
    \label{fig:bbcORD}
\end{figure}

\begin{figure}[t]
    \centering
    \begin{tikzpicture}
            \draw (0,0) node[inner sep=0]{\includegraphics[width=\linewidth ]{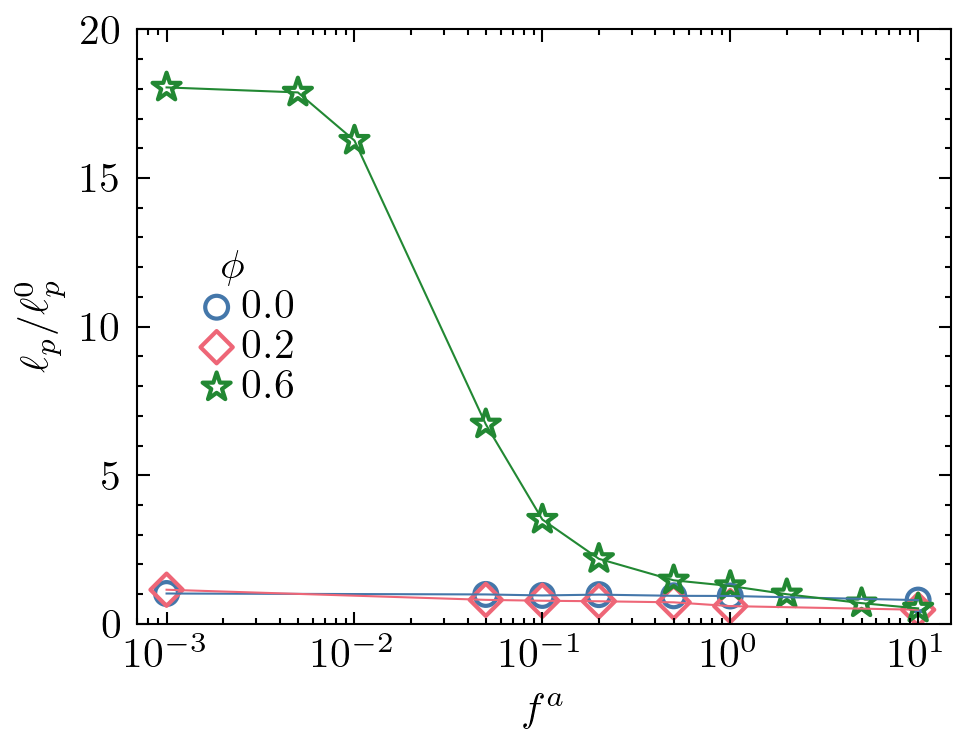}};
            \draw (0.5,3.3) node {{\large$\kappa=100$}};
    \end{tikzpicture}
    \caption{The effective persistence length normalized by the persistence length of passive chains in free surface $\ell_p/\ell_p^{0}$ against active force $f^a$ for various packing fraction $\phi$ and bending rigidity of $\kappa=100$.}
    \label{fig:LpORD}
\end{figure}
\subsection{Effect of periodic confinement on persistence length}
%The other conformational observable of interest is the bond correlation function and our main incentive here is to see how the confinements affect the general shape of active polymers. 
Having investigated the effects of periodic confinement on the end-to-end distance statistical features, next we look into its effect on the shape of polymers.  To this end, we compute the  bond-bond correlation function defined as $\langle cos(\theta(s))\rangle$, where $\theta(s)$ is the angle between two bond vectors with a curvilinear distance  $s$. Fig.~\ref{fig:bbcORD} shows the bond-bond correlation functions 
 for different   activity levels and bending rigidity values in free space and under strong confinement at $\phi=0.6$. 

For flexible polymers ($\kappa=1$), regardless of the degree of confinement  and activity, bond-bond correlations drop to zero at short curvilinear distances ($s<5$). The only notable effect at strong confinement ($\phi=0.6$) is the appearance of a weak negative dip  in $\langle cos(\theta(s))\rangle$ before decaying to zero reflecting the more shrunken conformation of polymers in caged state relative to the freely moving polymers. 
%This results in smaller conformation compared to same polymers without confinement and it is in agreement with the corresponding end-to-end distance results. \\

For stiff  chains ($\kappa=100$), strong confinement visibly leads to  
a slower decay of bond-bond correlation functions with curvilinear distance $s$, see  Figs.~\ref{fig:bbcORD}(c) and (d), the pace of which depends on the active force.
%have more persistent bond correlation along their contour as the plots suggest, however there exists visible differences between stiff chains with $\kappa=100$ in different conditions. At high packing fraction ($\phi=60$) and moderate activity ($f^a=0.1$), the chains maintain their backbone direction through longer contour distances $s$. 
%In order to compare  polymers with regards to this observation, 
To quantify this effect, we obtain the effective persistence lengths $\ell_p$ with the following protocol. First, we define $s_e$ as the  curvilinear distance at which the correlation becomes  equal to $e^{-1}$, \emph{i.e},  $\langle cos(\theta(s_e))\rangle \leq e^{-1}$. For the cases where the correlations never reach $e^{-1}$,  we choose $s_e=N-1$.  We then fit the bond correlation data in the range $0 \leq s \leq s_e$ by an exponential function $\langle cos(\theta(s))\rangle=\exp(-s/\ell_p)$ and  evaluate the persistence length. Fig.~\ref{fig:LpORD} shows the effective persistence length $\ell_p$ of stiff polymers with    $\kappa=100$ normalized by the  bare persistence length of passive free polymers, {\it i.e.}, $ \ell_p^0=2\kappa $ versus active force at three packing fractions. At $\phi=0.0$ and $\phi=0.2$, we observe no significant changes of $\ell_p$  with activity level. However, for the stronger  confinement at $\phi=0.6$, marked changes in the effective persistence  length emerge. 
%Similar to the trends observed for $\sqrt{\rangle R_e^2 \langle}$, persistence length  
As discussed earlier, the transport of stiff active chains within a dense periodic array of obstacles at  $\phi=0.6$  consists of a sequence  of straight paths within horizontal or vertical channels and channel switching events, the frequency of  which   increases with active force.  A slightly active stiff polymer with $f^a=10^{-3}$ and $\kappa=100$ almost always travels in the same channel,
and barely   switches to another one. Thus its backbone stays straight most of the time
resulting effectively in larger persistence length (almost 20 times larger than the  chains moving in free space. However, by increasing active force and thereby frequency of channel switching events, polymers bend more often. Hence, their effective persistence length decreases. 
%During channel switching events, polymers are bent and less extended, therefore 
%As previously discussed, by making the chains more active,  they are able to overcome the bending energy barrier associated with folding of the backbone. Hence, the frequency of channel switching events increases. 
  
 %The same narrative holds for less stiff polymers with $\kappa=10$. However, compared to chains with $\kappa=100$, the bending energy barriers for channel switching are lower, hence the difference  between  $\ell_p$ of free chains and those in tight confinement are less pronounced.

\section{Flexibility-dependent modes of transport}
%\subsection{Caging vs. channel switching}
In the previous section, we demonstrated that strong periodic confinement $\phi=0.6$  affects conformation of flexible and stiff active polymers differently. Increased confinement promotes greater contraction of  flexible polymers, whereas it results in more elongated polymer conformations for stiffer polymers. This difference leads to distinct modes of transport for flexible and stiff polymers within tight inter-obstacle pores. The motion of active flexible polymers in the periodic lattice of obstacles consists of a sequence of caging and hopping events, whereas the motion of stiff chains consists of travelling straight in the tight channels and occasional bending of polymers to switch  to an adjacent perpendicular channel.  
In this section, we examine the dynamics of individual active polymers  as they navigate through the interobstacle space by obtaining the distribution of the duration of caging events and  the time span of the directed motion in  channels.
%time spent being caged or unidirectional motion  and time needed to hop or switch  to another channel for flexible and stiff polymers.

\begin{figure}[t]
    \centering
    \begin{tikzpicture}
            \draw (0,0) node[inner sep=0]{\includegraphics[width=\linewidth ]{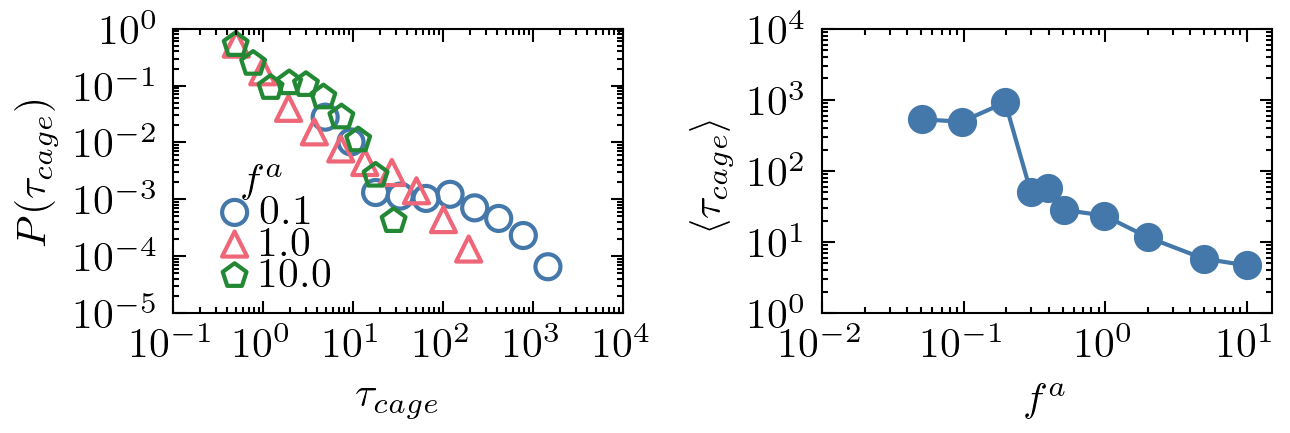}};
            
            \draw (-.7,1.) node {\textbf{(a)}};
            \draw (3.6,1.) node {\textbf{(b)}};
            %\draw (-.7,-.5) node {\textbf{(c)}};
            %\draw (3.6,-.5) node {\textbf{(d)}};
            
    \end{tikzpicture}
    \caption{a) The probability distribution function of caging time $\tau_{cage}$ for flexible chains with $\kappa=1$ in tight confinement with $\phi=0.6$ and for different activities. b) The mean values of  $\langle \tau_{cage} \rangle$. }%\sara{Please put minor ticks to make it  easier to read the values from the plots. Remove panel (e), you never discussed it in the text and it is not useful.}} 
    %e) The corresponding average caging period $T_{cage}$ as a function of $f^a$.   }
    \label{fig:caged}
\end{figure}

We first focus on flexible active polymers with $\kappa=1$. To be able to quantify the duration of caging events, we define a caged state when the end-to-end distance of a polymer is smaller than a threshold value $R^*$, which we choose to be the diameter of inter-obstacle pore $d_{cage}$, see Fig.~\ref{fig:Re_trapp} (a) and (b). For tight confinement with $\xi=2.5$, we have $d_{cage}\approx 10$ and  time intervals with  $R_e<R^*$ ($R_e>R^*$) characterize caging (hopping) events, see Figs.~\ref{fig:Re_trapp} (a) and (b).
%On the other hand  moments with $R_e > R_{cage}$ define polymer hopping. 
Fig.~\ref{fig:caged} (a)  shows the distribution of caging time  $\tau_{cage}$ for different active forces. We notice that the caging times span several orders of magnitude. The maximum caging time observed within our simulation time depends on the activity level  and is largest for the lowest active force.   
We do not recognize a power law behavior for distribution of caging time. %\sara{Mohammad, can you check if they follow $t^{\alpha}exp(-t/\tau_0) $?}.
From these distribution functions, we extract the mean duration of caging events $\langle \tau_{cage} \rangle$ as functions of active force as presented in  Fig.~\ref{fig:caged}(b). We find that the mean caging time is roughly constant up to $f^a=0.2$ and afterwards it decreases with $f^a$.  % $f^{a^{-0.65}}$
%for $f^a >0.2$. 

\begin{figure}[t]
    \centering
    \begin{tikzpicture}
            \draw (0,0) node[inner sep=0]{\includegraphics[width=\linewidth ]{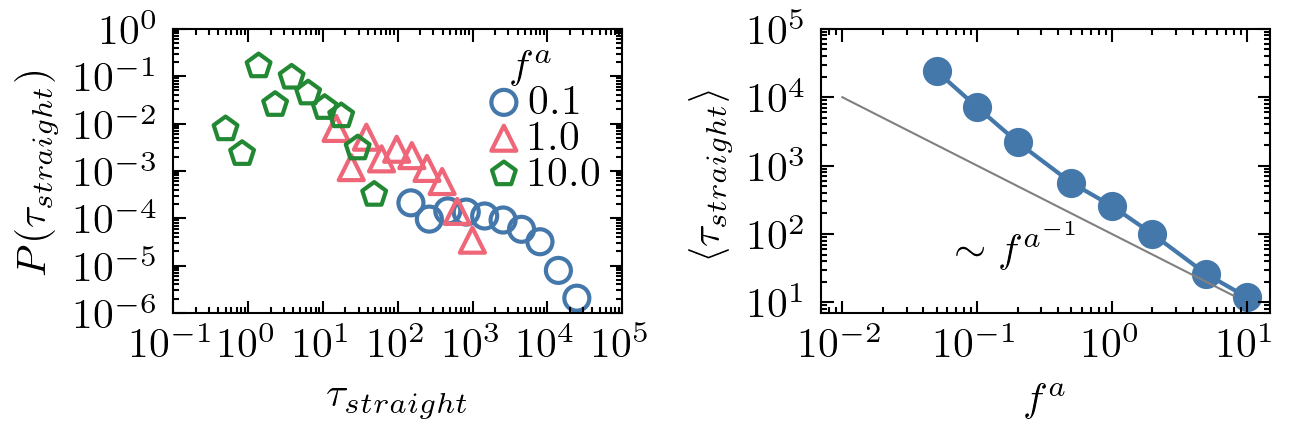}};
            %\draw (0,-4.8) node{\includegraphics[width=0.75\linewidth ]{Figures/T-CSwitch.png}};
            \draw (-2.8,-.3) node {\textbf{(a)}};
            \draw (1.5,-.3) node {\textbf{(b)}};
            %\draw (-2.9,-1.8) node {\textbf{(c)}};
            %\draw (1.5,-1.8) node {\textbf{(d)}};
            %\draw (-1.8,-5.4) node {\textbf{(e)}};
    \end{tikzpicture}
    \caption{a)The probability distribution function of straight traveling time $\tau_{straight}$ for polymers with $\kappa=100$ in tight confinement with $\phi=0.6$ and for different activities. b) The mean values of $\langle\tau_{straight}\rangle$.  }
    \label{fig:switch}
\end{figure}

For stiffer chains with $\kappa = 100$ in tight confinement, interactions of a polymer with periodic array of obstacles  result in a sequence of directed motion through  channels   and  channel switching events. Similar to the case of caging-hopping events, the distinction between unidirectional traveling and channel switching states   can be made by monitoring the instantaneous value of $R_e$ of individual polymers.
While travelling in a channel, the $R_e$ is comparable to the chain length $N$. However, during a channel switching event where a polymer bends, its $R_e$ decreases, see Fig.~\ref{fig:Re_trapp} (c) and (d).  To find the threshold, we refer to $P(R_e)$ presented in Fig.~\ref{fig:PDFRe}(d), where the probability distribution functions have a sharp peak at large $R_e$ corresponding to travelling within inter-obstacle channels. Among different activities, the sharp peak of $f^a=10$ is broader and includes a minimum value of $R_e\approx 90$. We therefore set the channel switching condition as $R_e \leq 90$ and find the time duration of  straight travel in channels $\tau_{straight}$ as the time interval for which $R_e > 90$. 
%define duration of a channel switching event $\tau_{switch}$ as the time interval for which $R_e < 90$. 
Fig.~\ref{fig:Re_trapp} (c) and (d) demonstrate how we distinguish between events of channel switching and straight motion  for stiff chains with $\kappa=100$.  In Fig.~\ref{fig:switch} (a), we present the PDF of  $\tau_{straight}$ for stiff chains with $\kappa=100$ at $\phi=0.6$. The typical time span of straight motion through the channels depends strongly on the  activity level.
The extracted  mean values $ \langle \tau_{straight} \rangle$ are presented in  Fig.~\ref{fig:switch}(b), which decrease with active force, scaling as $1/f^{a}$  for $f^a\gtrsim 0.1$.\\

This emergence of two different  scaling regimes in $ \langle \tau_{straight} \rangle$  against  active force can be understood in terms of competition between two time scales. The first one is the timescale of advection by active force for the frontal segment of active polar polymer and the second one is the timescale of the relaxation of bending fluctuations. 
A channel switching event involves traveling of a minimal segment of length $S$ of the polymer. $S$ represents the frontal segment of a polymer that is involved during the turning of the polymer as shown in Fig.~\ref{fig:TurnMech} when the head monomer moves from point $A$ to $B$. It can be approximated as $S\approx \pi/2(r_o + \xi/2)$.
 The time required for the head bead to travel a curvilinear distance $S$ is simply the advection time by active force $\tau_{a}^{adv}=S/f^a$. On ther hand the timescale for passive relaxation of bending fluctuation of a segment of curvilinear length $S$ of a semiflexible polymer, according to the wormlike chain model (WLC), is given by $\tau_{bend}=S^4/(2\kappa)$~\cite{transTime}. A channel switching event entails that the $\tau_{a}^{adv}< \tau_{bend}$ such that by increasing activity,  a chain can switch its channel before the  bending fluctuations on the scale of $S$ can relax. A quick calculation using $\xi=2.5$  for $\phi=0.6$ and $\kappa=100$ shows that $S\approx 15$ and the threshold activity (where $\tau_{a}^{adv}=\tau_{bend}$) is $f^a=0.06$. As can be seen for $f^a >  0.1$, $ \langle \tau_{straight} \rangle$ enters the $1/f^a$ scaling regime.

\iffalse In Fig.~\ref{fig:TurnMech}, we see snapshots of a stiff chain with $\kappa=100$ and $f^a=1.0$ switching its channel in a medium with $\phi=0.6$. The chain enters the crossing point denoted by A, travels a curvinlinear distance of $S$ while bending towards point B and leave the cross section in another channel. The time that it takes for the head bead to travel from A to B is simply the advection time $\tau_{S}^{adv}=S/f^a$, while according to wormlike chain model (WLC), the passive relaxation time of a bend on a segment with curvilinear length of $S$ is $\tau_{S}=S^4/(2\kappa)$ in free space~\cite{transTime}. By increasing activity, the advection time $\tau_{S}^{adv}$ becomes smaller than the bending relaxation time $\tau_{S}$, which means the chain can switch channel before the necessary bend relaxes. To find the threshold active force, we approximate the distance between A an B by $S\approx \sqrt{2}(r_o + \xi/2)$, then a quick calculation using $\xi=2.5$ ($\phi=0.6$) and $\kappa=100$ shows that for $f^a\geq 0.1$ the advection time becomes smaller than the bending relaxation time, therefore we expect the asymptotic $1/f^a$ behaviour of $ \langle \tau_{straight} \rangle$ to emerge above $f^a=0.1$.
%\sara{I do not understand how you exactly defined $\tau_{hop}$ and why it shows a non-monotonic behavior with activity!! I expect it to decrease with $f^a$.}   
\fi

\begin{figure}[t]
    \centering
    \begin{tikzpicture}
            \draw (0,0) node[inner sep=0]{\includegraphics[width=1\linewidth ]{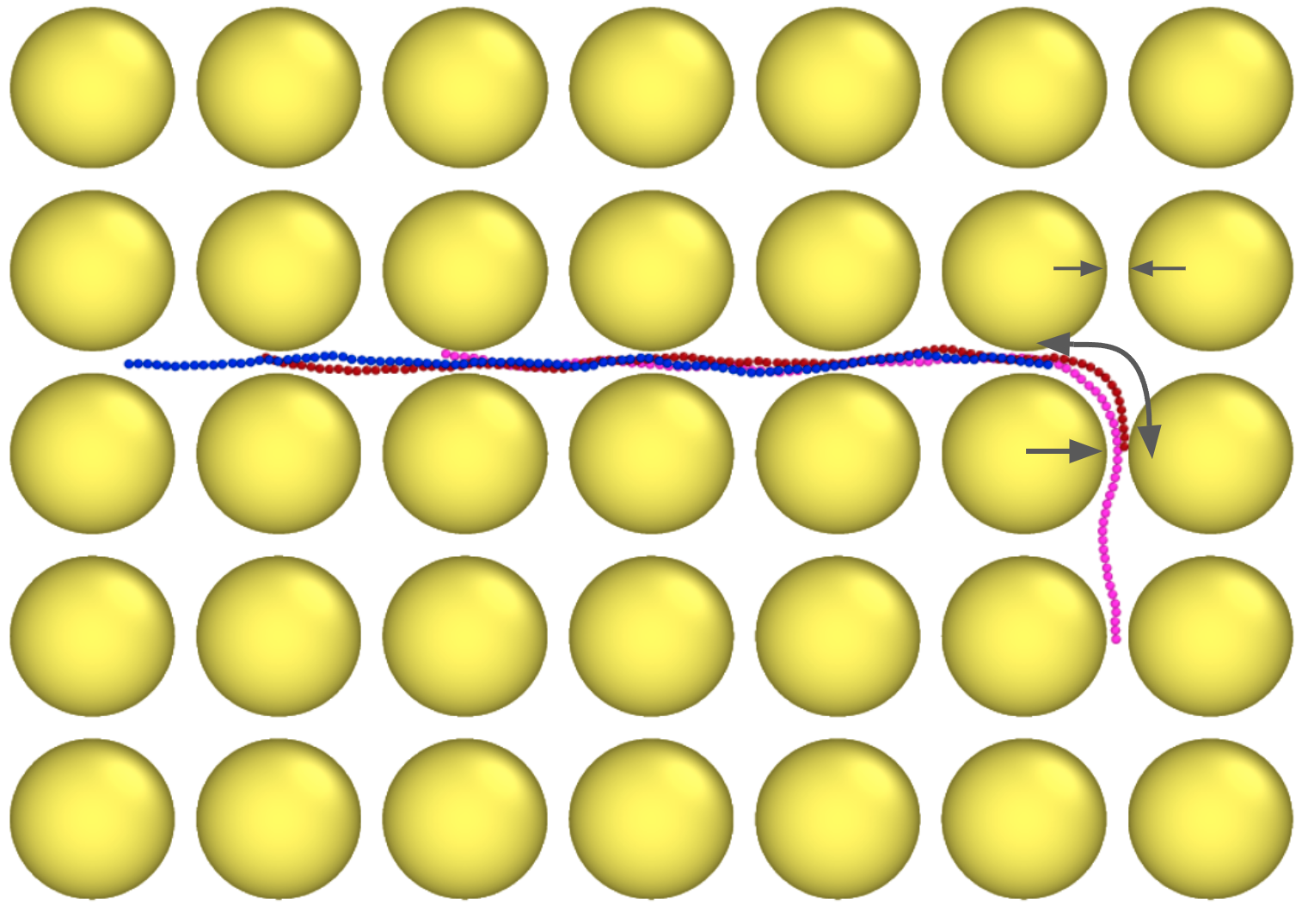}};
            \draw (2.48,1.1) node {\large{A}};
            \draw (3.5,0.1) node {\large{B}};
            \draw (2.7,-0.1) node {$r_o$};
            \draw (3.1,1.6) node {$\xi$};
            \draw (3.35,0.7) node {\large{$S$}};
    \end{tikzpicture}
    \caption{Snapshots of a chain with $\kappa=100$ and $f^a=1.0$ traveling in a periodic medium with $\phi=0.6$. The polymer enters the crossing area at point A. It actively changes the backbone orientation of its head segment 
 toward point B. The chain leaves the crossing in a new channel.
    }
    \label{fig:TurnMech}
\end{figure}
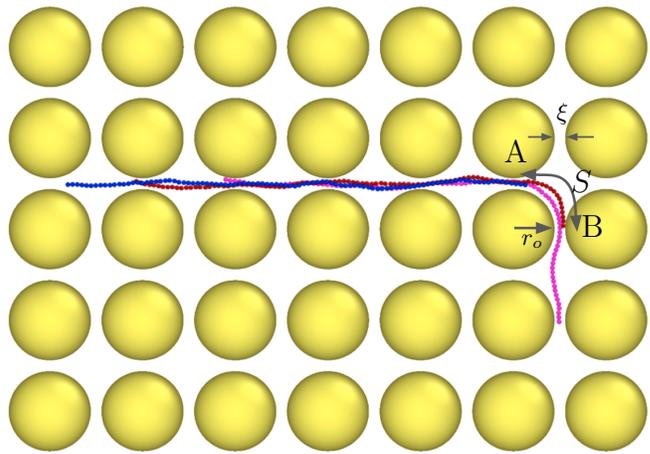

\iffalse
Similar to caging period, we define channel switching period $T_{switch}$ as the average time between two consecutive channel switching events. First, we calculate the channel switching probability $P_{switch}=\sum \tau_{switch}/T_{sim}$, where the summation is over all channel switching events. Then, noting that each complete channel switching takes an approximate time of $N/f^a$, we find $T_{switch}=\dfrac{N/f^a}{\sum \tau_{switch}/T_{sim}}$.\\

\begin{figure}[t]
    \centering
    \begin{tikzpicture}
            \draw (0,0) node[inner sep=0]{\includegraphics[width=1\linewidth ]{Figures/TURN_MECH.png}};
            \draw (-2.25,-3.5) node {Entering the crossing};
            \draw (2.25,-3.5) node {Leaving  the crossing};
            \draw (-1.35,1.7) node {\Large{A}};
            \draw (3.3,1.7) node {\Large{A}};
            \draw (1.75,2.5) node {\Large{B}};
            \draw (3.1,3.) node {\Large{C}};
    \end{tikzpicture}
    \caption{A channel switching event where on the left hand side the head bead enters a channel crossing at point A while being in a vertical channel. On the right hand side, the head bead collides with an obstacle at point B and leaves the crossing in a horizontal channel.}
    \label{fig:TurnMech}
\end{figure}
\fi

\section{Dynamical properties }
Having discussed the effects of periodic confinement on conformational properties of active polymers and their  modes of transport under tight confinement, we subsequently investigate the effects of periodic confinement on their dynamical properties. 
\subsection{Orientational dynamics\label{Sec:Orientattional}} \label{Sec:orientation}
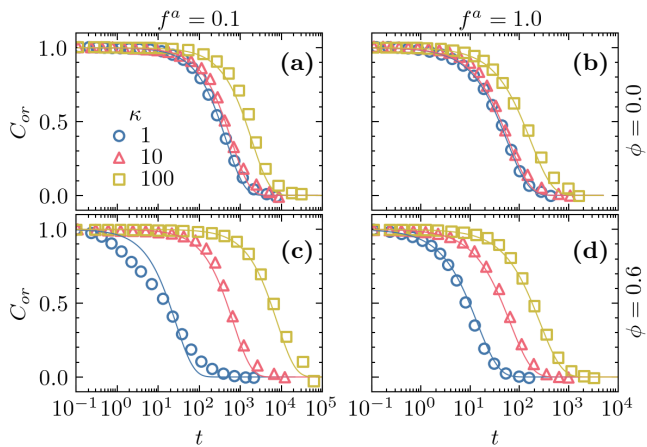
\begin{figure}[t]
    \centering
    \begin{tikzpicture}
            \draw (0,0) node[inner sep=0]{\includegraphics[width=\linewidth ]{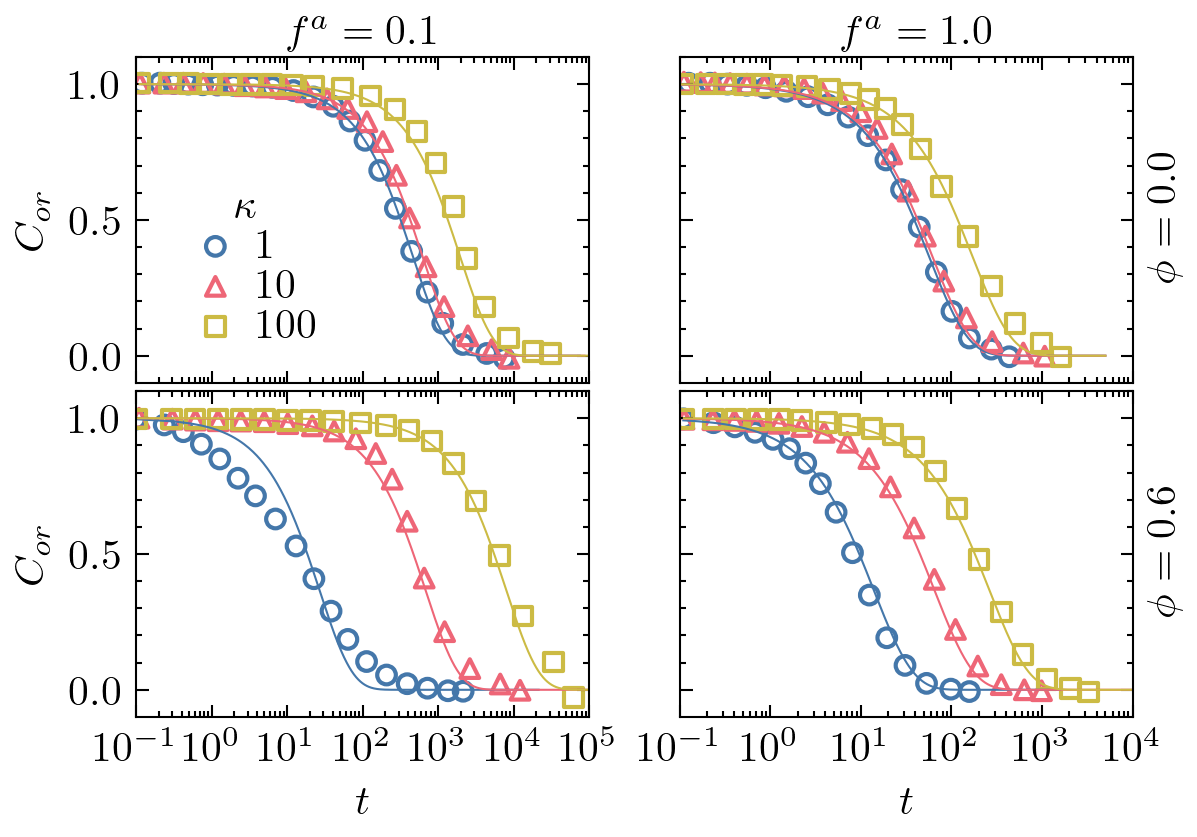}};
            \draw (-0.4,2.2) node {\textbf{(a)}};
            \draw (3.5,2.2) node {\textbf{(b)}};
            \draw (-0.4,-0.3) node {\textbf{(c)}};
            \draw (3.5,-0.3) node {\textbf{(d)}};
            
    \end{tikzpicture}
    \caption{The TACF of end-to-end unit vector, $C_{or}$, for different bending rigidity values $\kappa=1,10$ and 100 for free chains with active force a) $f^a=0.1$ and b) $f^a=1$ and in periodic lattice of obstacles with packing fraction of $\phi=0.6$ for active force c) $f^a=0.1$ and d) $f^a=1$. The lines depict exponential fits to the data for the time interval where $1\leq C_e(t) \leq e^{-1}$. }%\sara{please change $C_e$ to $C_{or}$. also, we can not read the begining of x-axis for the right panels. It is not definitely $10^5$. Could you please make another plot where you show results of $f^a=10$, instead of 1? This might be more instructive. I am not sure if this is the most useful presentation of data. Maybe better show the same active force and $\kappa$ but different $\phi$. }}
    \label{fig:CeORD}
\end{figure}
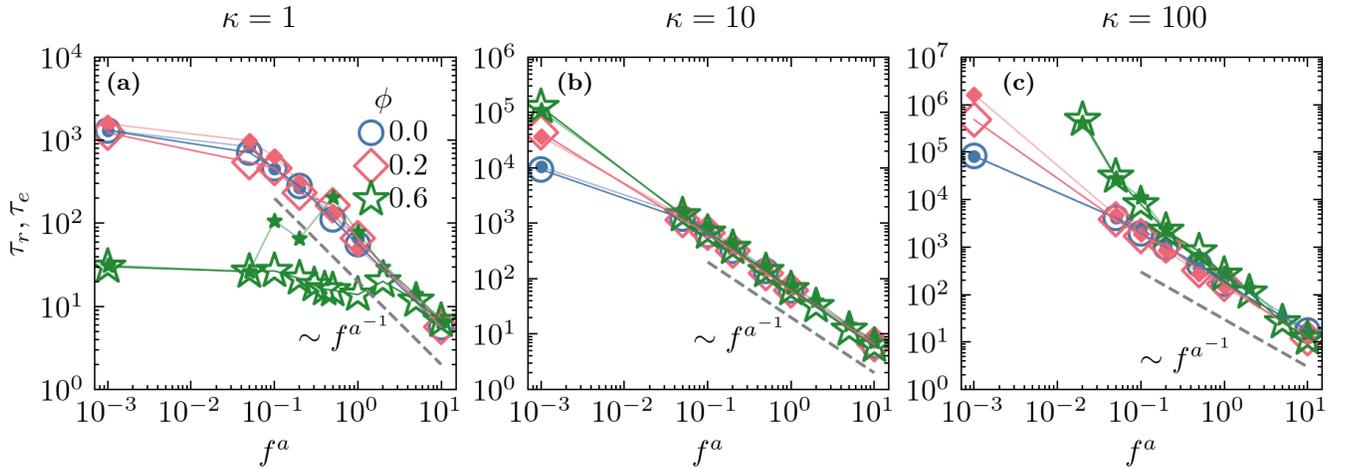
\begin{figure*}[t]
    \centering
    \begin{tikzpicture}
            \draw (0,0) node[inner sep=0]{\includegraphics[width=\linewidth ]{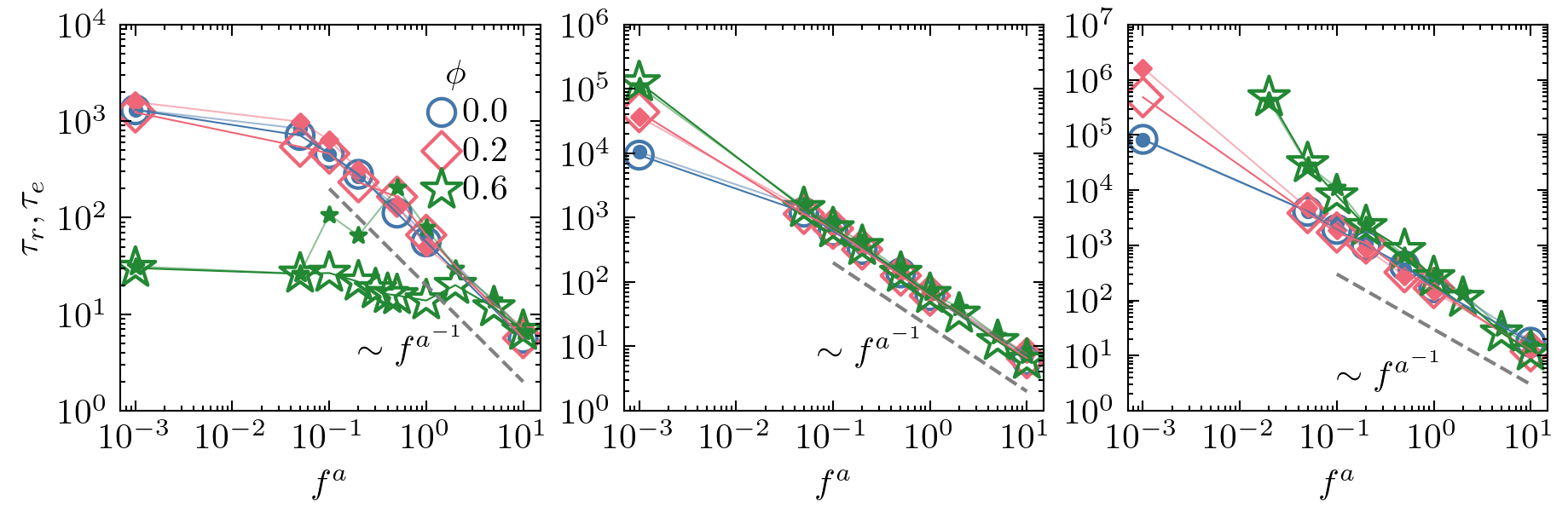}};
            \draw (-7.2,2.3) node {\textbf{(a)}};
            \draw (-1.2,2.3) node {\textbf{(b)}};    
            \draw (4.7,2.3) node {\textbf{(c)}};
            \draw (-5.4,3.2) node {{\large$\kappa=1$}};
            \draw (0.6,3.2) node {{\large$\kappa=10$}};
            \draw (6.5,3.2) node {{\large$\kappa=100$}};
    \end{tikzpicture}
    \caption{The end-to-end unit vector relaxation time $\tau_r$ (open symbols) versus active force $f^a$ for various packing fraction $\phi$ and bending rigidity of (a) $\kappa=1$, (b) $\kappa=10$ and c) $\kappa=100$. The closed symbols show the  relaxation time of the  end-to-end vector $\tau_e$ for $\kappa=10$ and  $\kappa=100$, respectively.} %\sara{do you have in the meantime, the $\tau_r$ of $\kappa=100$ for $f^a=0.01$? if not, please run the simulations. Could you also please add the $\tau_e$ data for $\kappa=1$, but make sure to compute directly $\langle R_e(t) \cdot R_e(0) \rangle$ and extract from it $\tau_e$, not from the normalized ACF.}}
    \label{fig:taueORD}
\end{figure*}
We start by examining the effects of periodic confinement on orientational dynamics of active polymers. Since a tangentially-driven polymer is polar (head-tail asymmetry), we define the end-to-end vector as $\vec{R}_e(t)=\vec{r}_{N}(t)-\vec{r}_{1}(t)$. 
%It can be easily shown that the net active force on center of mass is proportional to end-to-end vector, $\vec{F}^{a}(t)=\sum_{i=1}^{N}\vec{f}_{i}^a(t)=f^a \Vec{R}_{e}(t)/\ell$.
Thus, we characterize polymers orientational dynamics by the time auto-correlation function (TACF) of their end-to-end unit vector,
\begin{equation}
    C_{or}(t)= \langle \Vec{\hat{R}}_{e}(0).\Vec{\hat{R}}_{e}(t) \rangle.
    \label{eq:Ce}
\end{equation}
In Fig.~\ref{fig:CeORD}, we present the  orientational TACFs of active polymers in free space and in tight confinement $\phi=0.6$ as functions of lag time for different bending rigidity values and at two active forces $f^a=0.1$ and 1.0. In free space, for a given active force upon increase of bending stiffness, the orientational correlation functions decay slower, whereas by increasing active force, they decay faster, which is in agreement with prior reports for flexible active polymers~\cite{Bianco,Philipps_22,Fazelzadeh_23}. When introducing obstacles with $\phi=0.6$,   the periodic confinement can accelerate or slow down orientational dynamics depending on the bending stiffness and  activity level.  At  $f^a=0.1$  and $\phi=0.6$, we observe a much slower decay of the  orientational TACF  for stiff chains with $\kappa=100$ compared to the freely moving chains with the same active force and bending stiffness, whereas the orientational dynamics of flexible polymers becomes accelerated.  At higher active force of $f^a=1.0$, the orientational dynamics of flexible active polymers in tight periodic confinement is again faster than those of chains in free space. On the other hand, for very stiff polymers with $\kappa=100$, the orientational dynamics of active polymers in free space and tight periodic confinement become similar. 
%These results indicate  that at high activity levels the orientational dynamics of chains are predominantly governed by their activity and bending rigidity and large active forces can overcome the effects of periodic confinement.  \\

In order to quantify the effect of periodic confinement on the decay of  orientational dynamics, 
%have a quantitative understanding of how rapidly the orientational dynamics decays 
%as we vary the active force and degree of confinement,
we define a characteristic  reorientational relaxation time $\tau_r$ with the following protocol. First, we define $t_e$ as the shortest lag time at which the normalized correlation becomes less than $e^{-1}$, \emph{i.e},  $C_{or} \leq e^{-1}$. We then fit the data in the range $0 \leq t \leq t_e$ with an exponential function $C_{or}=\text{exp}(-t/\tau_r)$ from which we determine the reorientational relaxation time $\tau_r$. 
%It  was shown for tangentially driven chains that their end-to-end orientational dynamics become faster by increasing activity \sara{clarify under what conditions? free space or confined and what range of active force?}. Furthermore, whenever the orientational dynamics is majorly governed by activity the relaxation time scales as $\tau_e \sim 1/f^a$~~\cite{Bianco,ramirez}.\\
The extracted orientational relaxation times  as functions of  active force are shown with open symbols in Fig.~\ref{fig:taueORD}  for different values of bending rigidity $\kappa$ and $\phi$. First, we focus on the flexible polymer limit with $\kappa=1$ shown in  Fig.~\ref{fig:taueORD}(a). For free chains and those in moderate confinement with $\phi=0.2$,  $\tau_e$ scales as $1/f^a$ for $f^a \ge 0.1$, similar to the findings of prior research on isolated active polymers~\cite{Bianco,Fazelzadeh_23}. However, in tight confinement  $\phi=0.6$, chains with low and moderate active forces ($f^a \le 1$) have a much faster orientational dynamics and thus shorter relaxation times as a result of being caged between neighbouring obstacles. However, at higher activities $f^a \ge 2$, where the chains more frequently hop  to adjacent cages, the  orientational relaxation time becomes equal to that of freely moving active polymers. 
%the $1/f^a$ behaviour recovers for their relaxation time at $f^a \geq 0.5$.\\

For intermediate stiffness with $\kappa=10$, the orientational relaxation times are shown in Fig.~\ref{fig:taueORD}(b). For $f^a \ge 0.05$,  $\tau_r$ perfectly follows the $1/f^a$ scaling behaviour similar to isolated active flexible polymers~\cite{Bianco,Fazelzadeh_23},  suggesting the dominance of activity over confinement in this case. For stiff chains with $\kappa=100$,  free chains and those in moderate confinement with $\phi=0.2$ have identical relaxation times following the $1/f^a$ scaling for $f^a \ge 0.05$. In contrast, at $\phi=0.6$ we observe two different regimes. At higher activities ($f^a \ge 0.2$) the orientational relaxation times of strongly confined polymers are of the same order of magnitude as $\tau_r$ of isolated chains, while less active polymers have notably larger $\tau_r$. These results can be understood in view of confinement degree defined as the ratio of intrinsic persistence length of a polymer $\ell_p^0$ to the channel width of the medium $\xi$.  For strong confinement ($2\kappa/\xi =200/2.5 \gg 1$), the polymer is forced to keep travelling within one channel in a rather elongated conformation, see Fig.~\ref{fig:SnapsORD}. Hence, orientational dynamics is very slow as it entails overcoming the  bending energy of very stiff polymers for switching to another channel. At sufficiently high activities, the active force can overcome the bending energy barriers for channel switching and  the $ 1/f^a$ scaling behavior reemerges.
%Highly active chains can overcome the bending energy barriers for channel switching more easily, therefore the $\tau_e\sim 1/f^a$ recovers for them. 
However, less active chain can barely switch their channels, resulting in  orientational relaxation times almost one order of magnitude larger than that of isolated chains at $f^a=0.05$.
 
%As an example, in  Fig.~\ref{fig:taueORD}(c) at $f^a=0.05$, the relaxation time of  highly confined chains at $\phi=0.6$, $\tau_e=3\times 10^4$,  is almost 10 times larger than that of free chains $\tau_e=4\times 10^3$. 

\subsection{Translational dynamics}
\begin{figure}[b]
    \centering
    \begin{tikzpicture}
            \draw (0,0) node[inner sep=0]{\includegraphics[width=\linewidth ]{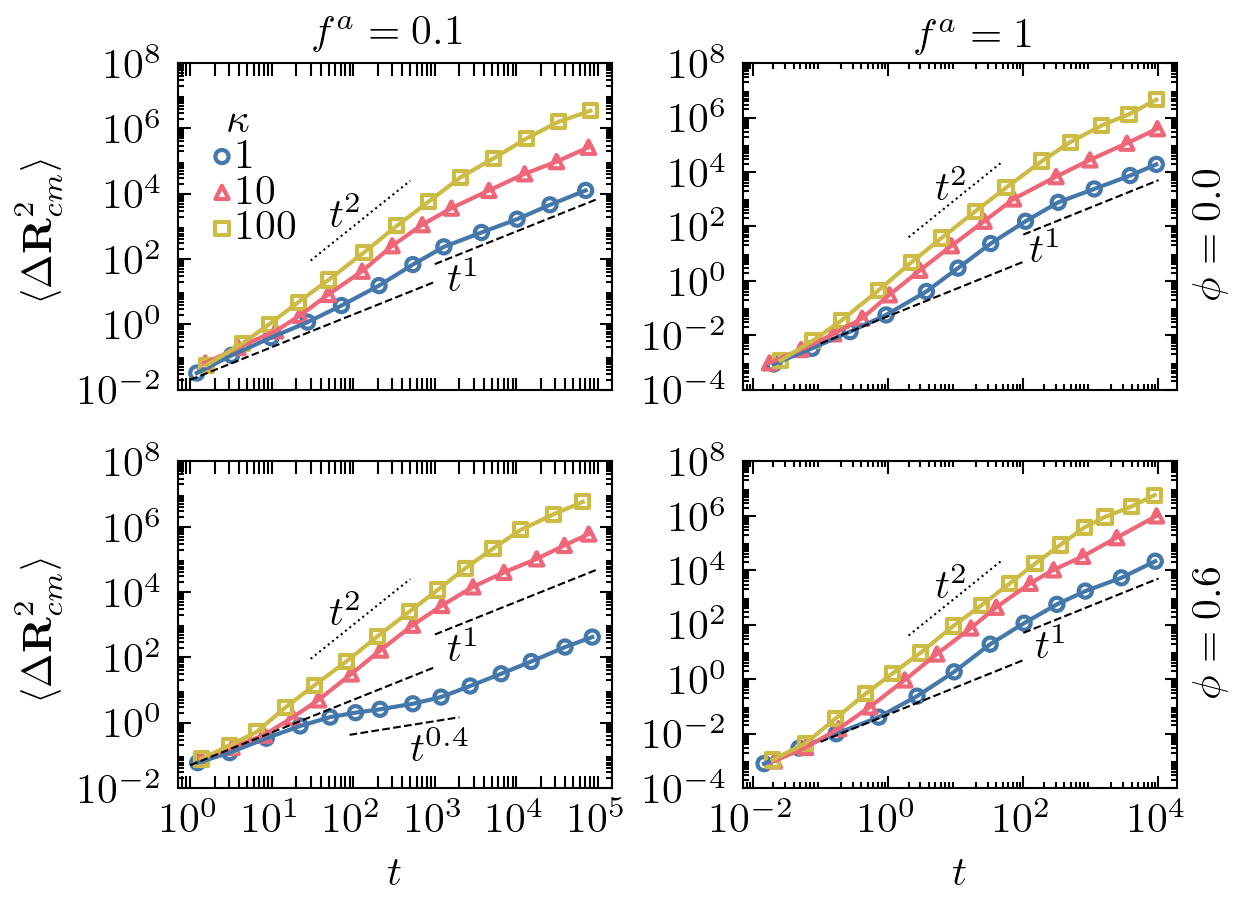}};
            \draw (-0.4,0.7) node {\textbf{(a)}};
            \draw (3.5,0.7) node {\textbf{(b)}};
            \draw (-0.4,-1.8) node {\textbf{(c)}};
            \draw (3.5,-1.8) node {\textbf{(d)}};
            
    \end{tikzpicture}
    \caption{The MSD $\Delta\vec{R}_{cm}^2$ for various bending rigidity values for free chains with active force a) $f^a=0.1$ and b) $f^a=1$ and at packing fraction of $\phi=0.6$ for active force c) $f^a=0.1$ and d) $f^a=1$. }
    \label{fig:msdORD}
\end{figure}
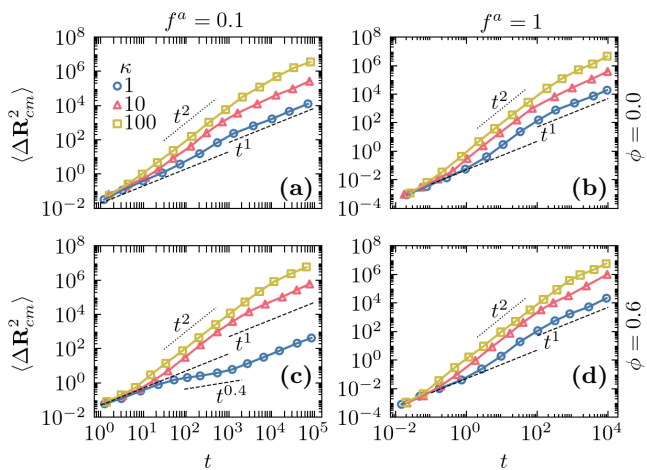

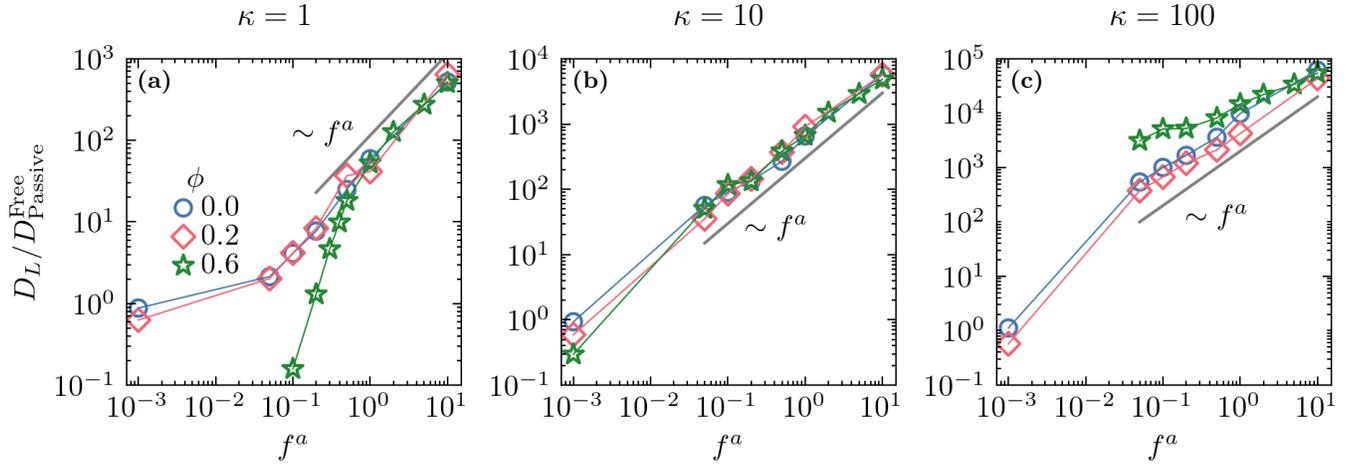
\begin{figure*}[t]
    \centering
   \begin{tikzpicture}
            \draw (0,0) node[inner sep=0]{\includegraphics[width=\linewidth ]{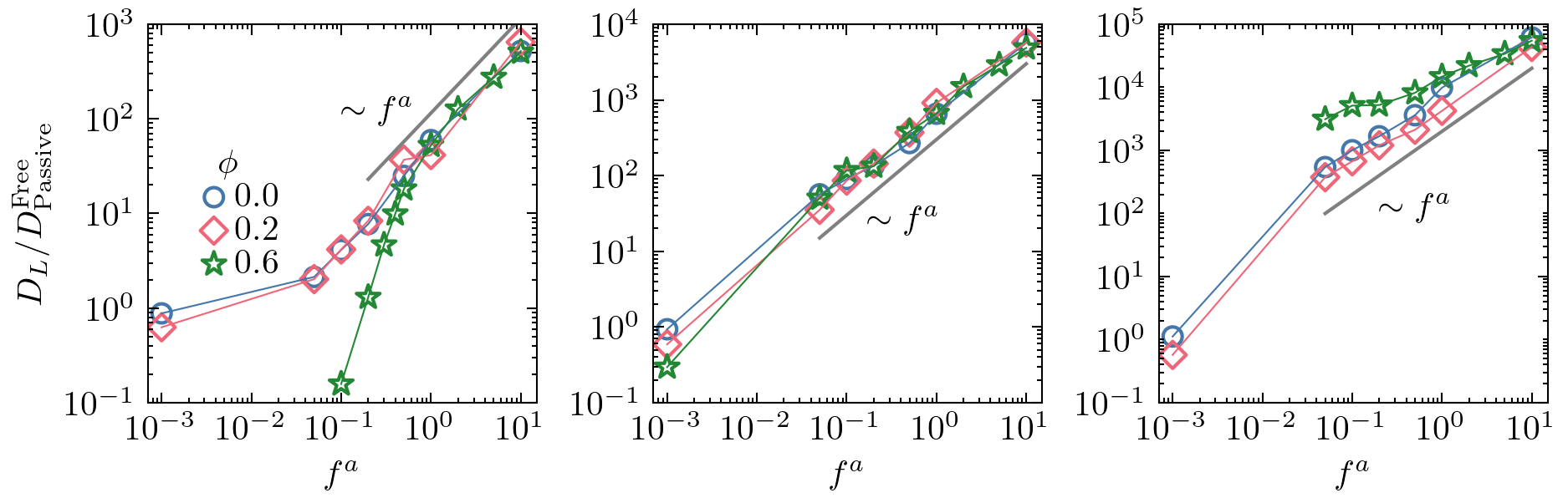}};
            \draw (-6.9,2.3) node {\textbf{(a)}};
            \draw (-1.1,2.3) node {\textbf{(b)}};    
            \draw (4.7,2.3) node {\textbf{(c)}};
            \draw (-5.3,3.2) node {{\large$\kappa=1$}};
            \draw (0.6,3.2) node {{\large$\kappa=10$}};
            \draw (6.5,3.2) node {{\large$\kappa=100$}};
    \end{tikzpicture}
    \caption{The normalized long time diffusion coefficient $D_L/D_{\text{Passive}}$ versus active force $f^a$ for  packing fractions $\phi=0$, 0.2 and 0.6 and bending rigidity of (a) $\kappa=1$, (b) $\kappa=10$ and (c) $\kappa=100$. In panel (a), for $\phi=0.6$, the chains with $f^a=0.001$ and $f^a=0.05$ never entered the final diffusive regime within our simulation time, therefore we cannot report   $D_L$ values for them. } %\sara{could you try to get longer data for $\kappa=100$ and $f^a=0.001, 0.05$?? and $\phi=0.6$, also $f^a= 0.05$?? and $\phi=0.2$ :: the data already exist for $f^a=0.05$, for smaller activity I ran the simulation, but they are still not long enough for dynamical properties calculation, such as $\tau_{r}$ }}
    \label{fig:DlORD}
\end{figure*}
Next we explore the translational dynamics of active polymers under periodic confinement by computing the mean squared displacement (MSD) of the center of mass. Defining the position of the center of mass at any time as $\vec{R}_{cm}(t)=\frac{1}{N}\sum \vec{r}_i(t)$, the MSD is computed as $\langle \Delta \vec{R}^2_{cm}(t)\rangle=\langle|\vec{R}_{cm}(t)-\vec{R}_{cm}(0)|^2\rangle$. Fig.~\ref{fig:msdORD} presents MSD curves as functions of lag time  in free space and strong confinement $\phi=0.6$ for different bending rigidity values and at two active forces $f^a=0.1$ and 1. 
%for various combinations of activities, packing fractions and bending rigidity values. \sara{This is a very vague and inaccurate way, as you only have 2 activities and two phi values.}
Regardless of defining parameters of chains and their surrounding media ($f^a$, $\kappa$ and $\phi$), in all the MSD curves, we observe three distinct scaling regimes.
At very short lag times, where the MSD values are initially proportional to $t$, the   fluctuations of random forces are more dominant over active force and interactions with obstacles, hence the translational motion is governed by thermal diffusion and the MSD has the form $\langle \Delta \vec{R}_{cm}^2\rangle=4D_{\text{Passive}}^{\text{Free}}t$, where $D_{\text{Passive}}^{\text{Free}}=D_{0}/N=0.01$ is the diffusion coefficient of a passive chain consisted of $N=100$ monomers moving in free space. At longer lag times active forces and interactions with obstacles come into play and  the chains enter an intermediate regime. 

At intermediate timescales, for chains in free space, this motion is ballistic  $\langle \Delta \vec{R}_{cm}^2 \rangle \sim t^2$  and is associated with the total active force on the center of mass of polymers~\cite{Bianco,ActivePolWinklerGompper,Fazelzadeh_23}. At this range of lag times, the motion of the center of mass of freely moving active polymers can roughly be interpreted as moving straight with a constant speed in the direction of total active force.  The net active force on center of mass for a tangentially-driven active polymer is given by $\vec{F}^{a}(t)=\sum_{i=1}^{N}\vec{f}_{i}^a(t)=f^a \Vec{R}_{e}(t)/\ell$,  being parallel to the end-to-end vector. Therefore, we expect that the timescale for departure from straight motion is set by  $\tau_r$, {\it i.e.} the relaxation time  of the end-to-end vector TACF. The observed ballistic regime at intermediate timescales remains intact even in the presence of obstacles for all the cases apart from low activity flexible chains with active force $f^a=0.1$ and under  tight confinement $\phi=0.6$, which  exhibit   a sub-diffusive behaviour with an MSD growing as $t^{0.4}$, see Fig.~\ref{fig:msdORD}(c).  The observed subdiffusive regime is a consequence of transient caging dynamics of flexible active polymers,  which are trapped in the inter-obstacle pores and from time to time hop to one of their adjacent cages.

At   longer lag times when $t \gg\tau_r$, after active chains have lost  the memory of their initial end-to-end vector direction,  a final diffusive regime with enhanced  long-time diffusion coefficient   $D_{L}$ emerges, {\it i.e.}  $\langle \Delta \vec{R}_{cm}^2 (t \gg \tau_e) \rangle  = 4D_{L}t$.   The $D_{L}$ values, extracted from linear fits of the MSD curves at large lag times,  against activity are presented in Fig.~\ref{fig:DlORD}  for $\phi=0$, 0.2 and 0.6 and different $\kappa$ values. For $\phi=0.6$, for $\kappa=1$ and 100 and $f^a \leq 0.05$, we did not observe a final diffusive regime within our simulation run time, therefore we cannot report $D_L$ values for these cases. 
For active chains in free space with $f^a >0.05$, the $D_L$ increases linearly with active force in agreement with theoretical predictions~\cite{Philipps_22} and previous simulations~\cite{Bianco,Fazelzadeh_23}.

A moderate periodic confinement  $\phi=0.2$ does not significantly affect the long-time diffusion coefficient for all values $\kappa$ and the linear scaling with $f^a$ remains intact. However, effect of tight confinement on $D_L$ very much depends on the bending rigidity. Interestingly,  the $D_L$ of active chains  with moderate bending stiffness  $\kappa=10$, see Fig.~\ref{fig:DlORD}(b) is not much affected by strong confinement  and it increases linearly with activity. At  $\phi=0.6$, the $D_L$ of flexible active polymers $\kappa=1$  for active forces $f^a<0.5$ is  lower than those of free chains, whereas $D_L$ of very stiff chains $\kappa=100$ for  active forces $ `0.05 \le f^a \le 0.5$ is remarkably enhanced. 
These contrasting trends reflect the different modes of transport for flexible   and stiff polymers. For flexible polymers, caging of polymers with low active forces slows down the diffusion as diffusion only occurs via occasional hopping events. In contrast, for stiff active polymers, the persistent unidirectional transport  in the inter-obstacle channels  helps them to diffuse through larger distances, thereby enhancing diffusion.
Nonetheless, at sufficiently high activity levels, the long-time diffusion coefficients  under tight confinement approach the $D_L$ of  active polymers in free space, regardless of their degree of flexibility. 

\subsection{Analytical calculations of center of mass dynamics}
%\subsection{Analytical estimate of long-time diffusion coefficient}
To rationalize the observed trends for the long-time diffusion coefficient of the  center of mass, we  derive  the equation of center of mass velocity $\vec{V}_{\text{cm}}(t)$  explicitly. By summing over all the monomers velocities described by Eq.~\eqref{eq:brownian}, we obtain:
\begin{equation}
   \gamma \vec{V}_{\text{cm}} = \frac{1}{N} (\vec{F}^a(t)+\vec{F}^{r}(t)  + \vec{F}^{o}(t) ),
 % \vspace{-2.5mm}
\end{equation}
where $\vec{F}^a=\sum_{i=1}^N \vec{f}_i^a(t)$ is the total active force, $\vec{F} ^r=\sum_{i=1}^N \vec{f}_i^r(t)$ is the sum of all the random forces with a zero mean and $\langle \vec{F}^r(t) \cdot \vec{F}^r (t') \rangle=4 N D_0 \gamma^2 \delta(t-t')$ and $\vec{F}^{o}(t)=\sum_{i=1}^N \nabla U_{excl}(\vec{r}_i)$ is the total force resulting from interactions with obstacles. As previously mentioned, for a tangentially driven polymer the total active force is proportional to end-to-end vector $\vec{F}^a(t) = f^a \vec{R}_{e}(t)/\ell$. 
%sara{You need to guide the reader and need a transition and explanation sentence before telling about further simplifications.}

In order to obtain the long-time diffusion, we need to compute the TACF of center of mass velocity. Taking into account that  correlations of other forces  with the total random force  vanish, the  TACF of the center of mass velocity is given by
\begin{eqnarray}
      C_v(t)&=&\langle\vec{V}_{cm}(t).\vec{V}_{cm}(0)\rangle  \label{eq:Cv}  \\ 
     &=& (\frac{f^a}{N \ell \gamma})^{2} \langle \vec{R}_e(t) \cdot \vec{R}_e(0) \rangle  \nonumber \\ &+& \dfrac{f^a}{N  \ell \gamma} [\langle \vec{R}_e(t)\cdot \vec{F}^{o}(0)+ \langle \vec{F}^{o}(t)   \cdot \vec{R}_e(0) \rangle ]\nonumber \\
     &+&  \dfrac{1}{N \gamma} \langle \vec{F}^{o}(t) \cdot \vec{F}^{o}(0) \rangle+
      (\dfrac{1}{N \gamma})^{2} \langle \vec{F}^{r}(t) \cdot \vec{F}^{r}(0)\rangle. \nonumber  
\end{eqnarray}
%It include contributions from end-to-end vector TACF, the correlation function of collision forces with obstacles and the end-to-end vector and TACFs of collisional and random forces. 

We can simplify the above equation for semiflexible polymers with $\kappa \ge 10$  taking into account the following considerations.  For $\kappa \ge 10$ and sufficiently large activities  $f^a\geq0.05$, the periodic confinement constrains  active polymers with $\kappa\geq 10$  to move in the inter-obstacle channels, so the frequency of collisions is low and interactions with obstacles are somewhat random.  
 Hence, we argue that  the contributions from correlations of end-to-end vector and collisional forces, {\it i.e.} the third line of Eq.~\eqref{eq:Cv}, are negligible. We emphasize that even though we have neglected these contributions, the effects of collisions with obstacles are reflected in the dynamics of the end-to-end vector.
 Collision events change  the mean end-to-end distance, see Fig.~\ref{fig:ReORD} as well as orientational dynamics, see  section~\ref{Sec:Orientattional}.  Especially, increasing the degree of confinement ($2\kappa/\xi$) significantly increases  orientational relaxation time for $\kappa=100$ and moderate active forces, see Fig.~\ref{fig:taueORD}(c).   

Assuming  that the fluctuations of end-to-end distance are negligible,   justified for stiff active polymers, see Fig.~\ref{fig:PDFRe} (b) and (d),   and  the TACF of end-to-end vector decays exponentially, we can approximate it as
\begin{equation}
    \langle \vec{R}_e(t) \cdot \vec{R}_e(0) \rangle =\langle \vec{R}_e(t)^2 \rangle e ^{\dfrac{-t}{\tau_e}}
\end{equation}
where the  decay time $\tau_e$ depends on activity, bending stiffness and packing fraction.  We have also presented $\tau_e$ as a function of active force for different values of $\kappa$ and $\phi$ in Fig.~\ref{fig:taueORD} as closed symbols. For $\kappa \ge 10$  $\tau_e$ and  $\tau_r$ (open symbols) values are in good agreement. In contrast for flexible polymers at $\phi=0.6$, where the polymers shrink in cages and extend during hopping $\tau_e$ and  $\tau_r$ are different at intermediate active forces. Hence, our proposed approximation for them are not valid.

\begin{figure}[t]
    \centering
    \begin{tikzpicture}
            \draw (0,0) node[inner sep=0]{\includegraphics[width=1\linewidth ]{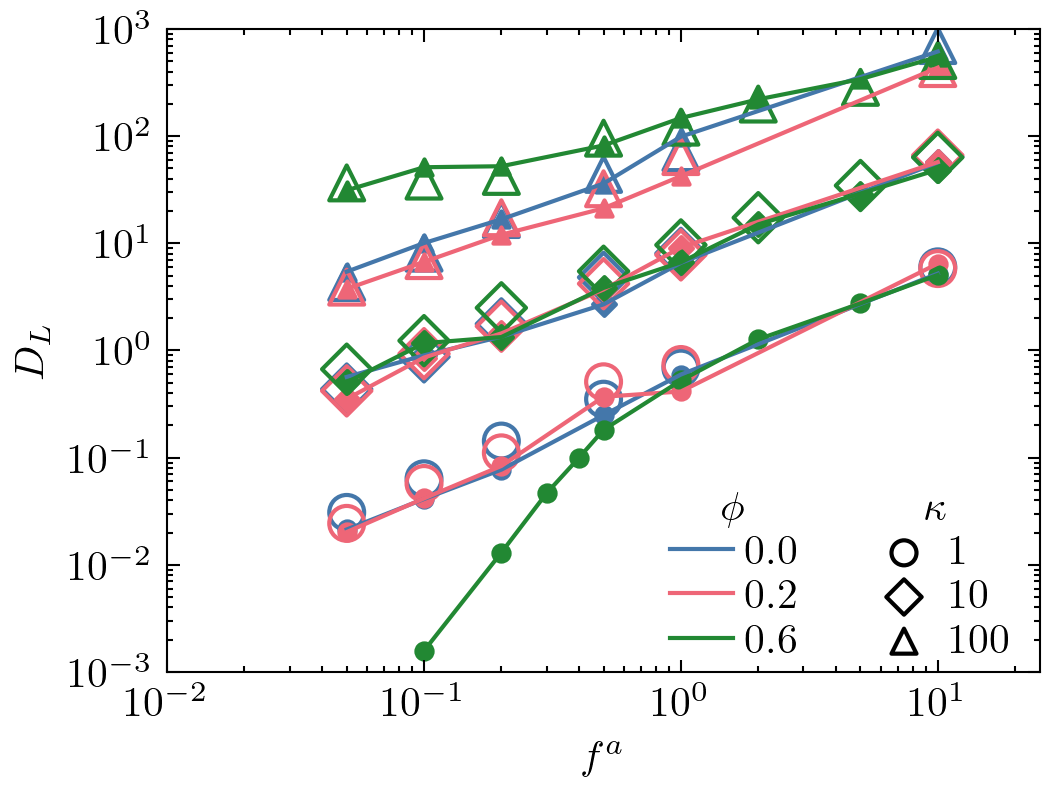}};
            %\draw (-0.25,1.7) node {\textbf{(a)}};
            
    \end{tikzpicture}
    \caption{Long-time diffusion coefficient  $D_L$ versus active force $f^a$ for different packing fraction for semiflexible polymers of bending stiffness $\kappa=1$,10 and 100. The closed symbols (with the lines as guides) show $D_L$ values extracted from MSD curves. The open symbols show the results of eq.~\eqref{eq:DLexplicit}. For the case of $\kappa=1$ and $\phi=0.6$ the results of Eq.~\eqref{eq:DLexplicit} are not presented as our assumptions are not valid in this case. }%\sara{more data is needed for $\kappa=1$ beyond $f^a=1$.}}
    \label{fig:DLTHEO}
\end{figure}

Taking into account the above considerations, we  obtain the  the following approximation for TACF of $\Vec{V}_{cm}$ of semiflexible polymers:
\begin{align}
    C_{v}(t>0) &\approx \dfrac{f^{a2}}{N^2 \ell^2 \gamma^2} \langle R_{e}^2\rangle \;e^{\dfrac{-t}{\tau_e}} \label{eq:Ce} \\
     &+ \dfrac{1}{N \gamma} \langle \vec{F}^{o}(t) \cdot \vec{F}^{o}(0) \rangle+(\dfrac{1}{N \gamma})^{2} \langle \vec{F}^{r}(t) \cdot \vec{F}^{r}(0) \rangle. \nonumber
\end{align}
Integrating this approximate $C_{v}$, we can obtain the long time diffusion coefficient as  $D_L=\frac{1}{2} \int_0^{\infty} dt   C_{v}(t)$. The terms in the second line of Eq.~\eqref{eq:Ce} includes contributions from TACFs of  collisional and random forces, which do not depend on the active force. Hence their contributions to the integral can be represented as a packing fraction and bending stiffness dependent  diffusion coefficient of a passive polymer $D_{\text{Passive}}^{\phi,\kappa}$. Consequently, $D_L$ can be approximated as 
\begin{equation}
    D_L \approx D_{\text{Passive}}^{\phi,\kappa} + \dfrac{f^{a2}\langle R_{e}^2\rangle \tau_e}{2N^2 \ell^2 \gamma^2}.
    \label{eq:DLexplicit}
\end{equation}
Passive stiff chains in high packing fraction media are confined to travel in inter-obstacle channels. Therefore their motion is effectively one dimensional with a diffusion coefficient of $D_{\text{Passive}}^{\text{Free}}/2$ \cite{ReptationofPassivePolymer}. By decreasing the degree of confinement the passive diffusion coefficient gradually increases until the $D_L$ of free chains is recovered. We calculate the predictions of Eq.~\eqref{eq:DLexplicit} using $\langle R_e^2 \rangle$ and $\tau_e$ from simulations. Comparison of $D_L$ extracted from MSD curves with results of Eq~\eqref{eq:DLexplicit}for semiflexible polymers, presented in Fig~.\ref{fig:DLTHEO}, show good agreement.  %There is however an exception, which is the case of flexible chains with $\kappa=1$ in tight confinement with $\phi=0.6$, where eq.~\eqref{eq:DLexplicit} fails to predict the correct diffusion coefficients since our assumptions for simplifying eq.~\eqref{eq:Cv} are not valid anymore.

\iffalse
\section{Discussion?}
Discussion of how $D_L$  can be interpreted based on analytical expression and interactions with obstacles
\fi

\section{Conclusion and Outlook}

We have computationally studied the effects of periodic confinement, created by a square lattice of circular obstacles, on conformational and dynamical properties of semiflexible tangentially driven active polymers. We considered two  packing fractions of obstacles, $\phi=0.2$ and 0.6. We find that effects of periodic confinement become significant only at the higher packing fraction $\phi=0.6$ and its impact on conformation and dynamics strongly depends on the degree of flexibility and the activity level. For flexible polymers, notable changes arise when polymer size becomes comparable to the pore size, whereas for semiflexible polymers confinement effects predominate when the persistence length is much larger than the pore size. Strong periodic confinement ($\phi=0.6$)  affects conformation of flexible and stiff  active polymers in distinct ways.\\

Flexible polymers in tight confinement become predominantly localized inside the  interobstacle cages, resulting in 
shrunken conformations.  However, active polymers occasionally succeed to hop from one cage to another via activity-induced conformational fluctuations enabling them to pass through the narrow channels in elongated conformations. This localization phenomenon is similar to what is found for transport of active polymers in disordered meida~\cite{Elham}.   Upon increase of active force, the time spent in cages decreases and frequency of hopping events increases, leading to more extended polymer conformations and larger mean end-to-end distances. As a result,    we observe a relatively sharp transition from a localized conformation to extended conformation as a function of active force. At low activity levels the orientational relaxation time of active flexible polymers is two orders of magnitudes smaller as the caged polymers rattle quickly inside the  confinement pores. The long-time diffusion coefficient of center of mass is also substantially reduced.  On the contrary, at high active forces polymers frequently hop from one cage to another. Interestingly,  both orientational relaxation time and the long-time diffusion coefficient of highly active polymers approach those in free space.

In the other  limit, strong confinement suppresses transverse fluctuations of stiff  semiflexible polymers and enables them  to migrate  ballistically in elongated conformations within  inter-obstacle channels.  As a result, at low activity levels, we observe a notable increase of orientational relaxation time  which  in turn leads to enhancement of long-time diffusion of center of mass due to enhancement of persistent motion. On the other hand, high  active forces can overcome the energetic costs associated with bending and enable  active polymers  to fold more frequently  and to switch from one   channel to another. This in turn results in decrease of orientational relaxation time and increase of long-time diffusion of center of mass  such that polymers at high active forces navigate in tightly confined periodic environment as quick as free active polymers.  

It is worth mentioning that despite the different modes of transport for flexible and stiff active polymers under tight confinement, at sufficiently large activity levels, we observe a robust scaling for the orientational relaxation time with active force decreasing as $1/f^a$. Moreover, the mean end-to-end distance at high activities  also approaches to that of free polymers regardless of degree of flexibility. 
We finally presented an analytical approach for the long-time diffusion coefficient of the center of the mass of stiffer polymers ($\kappa \geq 10$), using justified approximations. Our theoretical estimate is  in good agreement with the direct results of the simulations, providing insights into  roles of orientational relaxation time and mean end-to-end distance on the long-time diffusion coefficient. 

%Our results verify the robustness of the scaling behavior of the orientational relaxation times for tangentially driven polymers. We found out that at sufficiently large activity, regardless of the stiffness and the packing fraction of the medium, the orientational relaxation time of the chains decreases as $1/f^a$. 

%However, in tight confinements, we observed deviation from the scaling behaviour for less active chains ($f^a<1$), specially for very flexible ($\kappa=1$) and very stiff chains ($\kappa=100$).\\

In summary, the general emerging pattern observed in the  dynamics of active chains with varying degrees of flexibility is that the transport  of highly active chains remains unaffected by the level of confinement. On the contrary, at low activity levels,   flexibility degree of active polymers plays a critical role for their transport through ordered porous media. This study suggests an optimal degree of flexibility for migration of active deformable particles in tight periodic confinement. This study is only the first step in understanding the motion of semiflexible tangentially driven polymers in heterogeneous ordered media and it calls for further investigations  on the role of chain length, type of periodic lattice and generalization to   periodic and disordered three-dimensional media in the future.

%this study is only the first step in understanding the motion of semiflexible tangentially driven polymers in heterogeneous media. By randomly dislocating the obstacles from their lattice positions up to a threshold distance, one can control the degree of disorder in the medium and investigate the effects of disorder on the polymers. 
%hydrodynamic effects.
%%%%%%%%%
\section{Supplementary material}
The supplementary material supports the results presented in
the main text and contains movie files.
\section{Acknowledgement}
We acknowledge A. Deblais and R. Sinaasapple for fruitful discussions. The computations were carried out on the Dutch National e-Infrastructure with the support of the SURF Cooperative. This work was part of the D-ITP consortium, a program of the Netherlands Organization for Scientific Research (NWO) that is funded by the Dutch Ministry of Education, Culture and Science (OCW).% Z. Mokhtari would like to acknowledge Germany’s Excellence Strategy – MATH+ : The Berlin Mathematics Research Center (EXC-2046/1) - project ID:  390685689 (subproject EF4-10) for partial support of this project.
%%%%%%%%

 \nocite{HOOMDBLUE}
\bibliography{active_polymer.bib}

\end{document}

\title[]{Directing motion of active semiflexible polymers in periodic array of obstacles\\ Supplementary Material}
\maketitle

\section{Dimensionless groups and simulation details}

The simulations are carried out in  HOOMD-BLUE~\cite{HOOMDBLUE} package  with home-made modification to include the active force using a GPU implementation.   The equations of motion  are solved using a velocity-Verlet scheme.  
 To produce the desired statistics, for each set of parameters \emph{i.e}, $(f^a,\kappa,\phi)$  we have simulated 120 independent chain configurations. For this purpose, we first equilibrated passive polymers of different chain lengths and then we used them as starting configurations for active polymers. The simulations consist of $N_{run}$ timesteps until the polymers reach a dynamical steady state, to ensure that  no vestige of the initial conditions are remained.
 Subsequently they are followed by a production run of $N_{pro}$ timesteps to compute the structural and dynamical observable quantities. We  verified that the total   time in the production run $t_{pro}=n_{pro}* dt$  is at least  four times larger than the relaxation time of the end-to-end vector $\tau_{r}$.  
 A summary of   simulation  parameters, including timestep $dt$ can be found in Table ~\ref{tableSIMSUM}.
\begin{table}[h]
\begin{tabular}{|c|c|c|c|c|c|c|}
\hline
  $\;f^{a}\;$ & $\kappa$ & $\phi$ &  $\;N_{run}\;$ & $\;N_{pro}\;$ & $\;dt\;$\\
\hline
     0.001 & 1 & 0 & 1e8 & 5e8 & 1e-4\\ 
\hline
0.0 & 1 & 0.001 & 1e8 & 5e8 & 1e-4\\ \hline
0.0 & 1 & 0.05 & 1e8 & 5e8 & 1e-4 \\ \hline
0.0 & 1 & 0.1 & 1e8 & 1e8 & 1e-4 \\ \hline
0.0 & 1 & 0.2 & 1e8 & 1e8 & 1e-4 \\ \hline
0.0 & 1 & 0.5 & 1e8 & 1e8 & 1e-4 \\ \hline
0.0 & 1 & 1.0 & 1e8 & 1e8 & 1e-4 \\ \hline
0.0 & 1 & 10.0 & 1e7 & 1e7 & 1e-4 \\ \hline

0.0 & 10 & 0.001 & 1e8 & 5e8 & 1e-4 \\ \hline
0.0 & 10 & 0.05 & 1e8 & 5e8 & 1e-4 \\ \hline
0.0 & 10 & 0.1 & 1e8 & 1e8 & 1e-4 \\ \hline
0.0 & 10 & 0.2 & 1e8 & 1e8 & 1e-4 \\ \hline
0.0 & 10 & 0.5 & 1e8 & 1e8 & 1e-4 \\ \hline
0.0 & 10 & 1.0 & 1e8 & 1e8 & 1e-4 \\ \hline
0.0 & 10 & 10.0 & 1e7 & 1e7 & 1e-4 \\ \hline

0.0 & 100 & 0.001 & 1e8 & 5e8 & 1e-4 \\ \hline
0.0 & 100 & 0.05 & 1e8 & 5e8 & 1e-4 \\ \hline
0.0 & 100 & 0.1 & 1e8 & 1e8 & 1e-4 \\ \hline
0.0 & 100 & 0.2 & 1e8 & 1e8 & 1e-4 \\ \hline
0.0 & 100 & 0.5 & 1e8 & 1e8& 1e-4 \\ \hline
0.0 & 100 & 1.0 & 1e8 & 1e8& 1e-4 \\ \hline
0.0 & 100 & 10.0 & 1e7 & 5e7 & 1e-4 \\ \hline

0.2 & 1 & 0.001 & 1e8 & 5e8 & 1e-4 \\ \hline
0.2 & 1 & 0.05 & 1e8 & 5e8 & 1e-4 \\ \hline
0.2 & 1 & 0.1 & 1e8 & 1e9 & 1e-4 \\ \hline
0.2 & 1 & 0.2 & 1e8 & 4e8 & 1e-4 \\ \hline
0.2 & 1 & 0.5 & 1e7 & 2e7 & 1e-4 \\ \hline
0.2 & 1 & 1.0 & 1e7 & 1e7 & 1e-4 \\ \hline
0.2 & 1 & 10.0 & 1e7 & 1e7 & 1e-4 \\ \hline

0.2 & 10 & 0.001 & 1e8 & 5e8 & 1e-4 \\ \hline
0.2 & 10 & 0.05 & 1e8 & 5e8 & 1e-4 \\ \hline
0.2 & 10 & 0.1 & 1e8 & 5e8 & 1e-4 \\ \hline
0.2 & 10 & 0.2 & 1e8 & 4e8 & 1e-4 \\ \hline
0.2 & 10 & 0.5 & 1e7 & 1e7 & 1e-4 \\ \hline
0.2 & 10 & 1.0 & 1e7 & 1e7 & 1e-4 \\ \hline
0.2 & 10 & 10.0 & 1e7 & 1e7 & 1e-4 \\ \hline
0.2 & 100 & 0.001 & 1e8 & 5e8 & 1e-4 \\ \hline

\end{tabular}
\hfill
\begin{tabular}{|c|c|c|c|c|c|c|}
\hline
  $\;f^{a}\;$ & $\kappa$ & $\phi$ &  $\;N_{run}\;$ & $\;N_{pro}\;$ & $\;dt\;$\\
\hline
0.2 & 100 & 0.05 & 1e8 & 4e8 & 1e-4 \\ \hline

0.2 & 100 & 0.1 & 1e8 & 1e9 & 1e-4 \\ \hline
0.2 & 100 & 0.2 & 1e8 & 4e8 & 1e-4 \\ \hline
0.2 & 100 & 0.5 & 1e7 & 1e7 & 1e-4 \\ \hline
0.2 & 100 & 1.0 & 1e7 & 1e7 & 1e-4 \\ \hline
0.2 & 100 & 10.0 & 1e7 & 1e7& 1e-4 \\ \hline

0.6 & 1 & 0.001 & 1e8 & 5e8 & 1e-4 \\ \hline
0.6 & 1 & 0.05 & 1e8 & 4e8 & 1e-4 \\ \hline
0.6 & 1 & 0.1 & 1e8 & 1e9 & 1e-4 \\ \hline
0.6 & 1 & 0.2 & 1e8 & 4e8 & 1e-4 \\ \hline
0.6 & 1 & 0.3 & 1e8 & 1e8 & 1e-4 \\ \hline
0.6 & 1 & 0.4 & 1e8 & 1e8 & 1e-4 \\ \hline
0.6 & 1 & 0.5 & 1e7 & 1e7 & 1e-4 \\ \hline
0.6 & 1 & 1.0 & 1e7 & 1e7 & 1e-4 \\ \hline
0.6 & 1 & 2.0 & 1e7 & 1e7 & 1e-4 \\ \hline
0.6 & 1 & 5.0 & 1e7 & 1e7 & 1e-4 \\ \hline
0.6 & 1 & 10.0 & 1e7 & 1e7 & 1e-4 \\ \hline

0.6 & 10 & 0.001 & 1e8 & 5e8 & 1e-4 \\ \hline
0.6 & 10 & 0.05 & 1e8 & 4e8 & 1e-4 \\ \hline
0.6 & 10 & 0.1 & 1e8 & 1e9 & 1e-4 \\ \hline
0.6 & 10 & 0.2 & 1e8 & 4e8 & 1e-4 \\ \hline
0.6 & 10 & 0.5 & 1e8 & 4e8 & 1e-4 \\ \hline
0.6 & 10 & 1.0 & 1e7 & 1e8 & 1e-4 \\ \hline
0.6 & 10 & 2.0 & 1e7 & 1e8 & 1e-4 \\ \hline
0.6 & 10 & 5.0 & 1e7 & 1e8 & 1e-4 \\ \hline
0.6 & 10 & 10.0 & 1e7 & 1e7 & 1e-4 \\ \hline

0.6 & 100 & 0.001 & 1e8 & 2e9 & 1e-4 \\ \hline
0.6 & 100 & 0.005 & 1e8 & 2e9 & 1e-4 \\ \hline
0.6 & 100 & 0.01 & 1e8 & 2e9 & 1e-4 \\ \hline
0.6 & 100 & 0.05 & 1e8 & 2e9 & 1e-4 \\ \hline
0.6 & 100 & 0.1 & 1e8 & 2e9 & 1e-4 \\ \hline
0.6 & 100 & 0.2 & 1e8 & 4e8 & 1e-4 \\ \hline
0.6 & 100 & 0.5 & 1e8 & 1e8 & 1e-4 \\ \hline
0.6 & 100 & 1.0 & 1e8 & 1e8 & 1e-4 \\ \hline
0.6 & 100 & 2.0 & 1e7 & 2e7 & 1e-4 \\ \hline
0.6 & 100 & 5.0 & 1e7 & 2e7 & 1e-4 \\ \hline
0.6 & 100 & 10.0 & 1e7 & 1e7 & 1e-4 \\ \hline
\end{tabular}

\caption{\label{tableSIMSUM}Summary of simulation for chains of various activities and bending stiffness values in free space, low and high confinement. The tables contains the information about the number of thermalizing steps $N_{run}$ before the actual simulation and then the total production steps of $N_{pro}$ and the corresponding step size $dt$.}
\end{table}
\section{Description of videos of simulation}
 We provide videos of the motion of active polymers in different media. The descriptions of the videos are included in Table.~\ref{tab:video}.
 \begin{table}[h]
     \centering
     \begin{tabular}{|c|c|c|c|c|c|}
     \hline
          Video & $\phi$ & $\kappa$ & $f^a$ & $dt$ & $N_{frame}$  \\ \hline 
          S1.mp4 & 0.0 & 1 & 1.0 & $10^{-4} $ & 1  \\ \hline
$\:\:$S2.mp4$\:\:$ & 0.0 & 10 & 1.0 & $10^{-4} $ & 5e3  \\ \hline
$\:\:$S3.mp4 $\:\:$& 0.0 & 100 & 1.0 & $10^{-4} $ & 5e3  \\ \hline
$\:\:$S4.mp4 $\:\:$& 0.2 & 1 & 1.0 & $10^{-4} $ & 5e3  \\ \hline
$\:\:$S5.mp4 $\:\:$& 0.2 & 10 & 1.0 & $10^{-4} $ & 5e3  \\ \hline
$\:\:$S6.mp4$\:\:$ & 0.2 & 100 & 1.0 & $10^{-4} $ & 5e3  \\ \hline
$\:\:$S7.mp4$\:\:$ & 0.6 & 1 & 1.0 & $10^{-4} $ & 5e3  \\ \hline
$\:\:$S8.mp4$\:\:$ & 0.6 & 10 & 1.0 & $10^{-4} $ & 5e3  \\ \hline
$\:\:$S9.mp4$\:\:$ & 0.6 & 100 & 1.0 & $10^{-4} $ & 5e3 \\ \hline
$\:\:$S10.mp4$\:\:$ & 0.6 & 1 & 0.1 & $10^{-4} $ & 5e4  \\ \hline
$\:\:$S11.mp4$\:\:$& 0.6 & 1 & 10.0 & $10^{-4} $ & 5e3  \\ \hline
     \end{tabular}
     \caption{Description of videos. The number of time steps between frames is denoted by $N_{frame}$.}
     \label{tab:video}
 \end{table}

\bibliography{active_polymer.bib}